\documentstyle[11pt,epsf]{article}

\def \tr{{\rm tr}}

\def \det{{\rm det}}

\def \log {{\rm log}}

\def \wt {\widetilde}

\def \be {begin {equation}}
\def \ee {end{equation}}
\def\tilde{\widetilde}
\def\bar{\overline}
\def\hat{\widehat}
\def\*{\star}
\def\({\left(}		\def\BL{\Bigr(}
\def\){\right)}		\def\BR{\Bigr)}
\def\[{\left[}		
\def\]{\right]}

\def\frac#1#2{{#1 \over #2}}		
\def\inv#1{{1 \over #1}}
\def\half{{1 \over 2}}

\def\2pi{\hbox{$2\pi i$}}

\def\dsl{\raise.15ex\hbox{/}\kern-.57em\partial}
\def\Dsl{\,\raise.15ex\hbox{/}\mkern-.13.5mu D}
\def\be{\beta}

\def\ep{\epsilon}
\def\la{\lambda}

\font\numbers=cmss12
\font\upright=cmu10 scaled\magstep1
\def\stroke{\vrule height8pt width0.4pt depth-0.1pt}
\def\topfleck{\vrule height8pt width0.5pt depth-5.9pt}
\def\botfleck{\vrule height2pt width0.5pt depth0.1pt}
\def\Zmath{\vcenter{\hbox{\numbers\rlap{\rlap{Z}\kern 0.8pt\topfleck}\kern
2.2pt
                   \rlap Z\kern 6pt\botfleck\kern 1pt}}}
\def\Qmath{\vcenter{\hbox{\upright\rlap{\rlap{Q}\kern
                   3.8pt\stroke}\phantom{Q}}}}
\def\Nmath{\vcenter{\hbox{\upright\rlap{I}\kern 1.7pt N}}}
\def\Cmath{\vcenter{\hbox{\upright\rlap{\rlap{C}\kern
                   3.8pt\stroke}\phantom{C}}}}
\def\Rmath{\vcenter{\hbox{\upright\rlap{I}\kern 1.7pt R}}}
\def\Z{\ifmmode\Zmath\else$\Zmath$\fi}
\def\Q{\ifmmode\Qmath\else$\Qmath$\fi}
\def\N{\ifmmode\Nmath\else$\Nmath$\fi}
\def\C{\ifmmode\Cmath\else$\Cmath$\fi}
\def\R{\ifmmode\Rmath\else$\Rmath$\fi}

\def\be{ \begin{eqnarray} }
\def\ee{ \end{eqnarray} }
\def\non { \nonumber\\ }

\def\bea{\begin{eqnarray}}
\def\eea{\end{eqnarray}}

\def\beq{\begin{equation}}
\def\eeq{\end{equation}}
\def\ba{\beq\new\begin{array}{c}}
\def\ea{\end{array}\eeq}
\def\be{\ba}
\def\ee{\ea}

\def\Tr{{\rm Tr}}

\makeatletter
\newdimen\normalarrayskip              
\newdimen\minarrayskip                 
\normalarrayskip\baselineskip
\minarrayskip\jot
\newif\ifold             \oldtrue            \def\new{\oldfalse}
\def\arraymode{\ifold\relax\else\displaystyle\fi} 
\def\eqnumphantom{\phantom{(\theequation)}}     
\def\@arrayskip{\ifold\baselineskip\z@\lineskip\z@
     \else
     \baselineskip\minarrayskip\lineskip2\minarrayskip\fi}
\def\@arrayclassz{\ifcase \@lastchclass \@acolampacol \or
\@ampacol \or \or \or \@addamp \or
   \@acolampacol \or \@firstampfalse \@acol \fi
\edef\@preamble{\@preamble
  \ifcase \@chnum
     \hfil$\relax\arraymode\@sharp$\hfil
     \or $\relax\arraymode\@sharp$\hfil
     \or \hfil$\relax\arraymode\@sharp$\fi}}
\def\@array[#1]#2{\setbox\@arstrutbox=\hbox{\vrule
     height\arraystretch \ht\strutbox
     depth\arraystretch \dp\strutbox
     width\z@}\@mkpream{#2}\edef\@preamble{\halign
\noexpand\@halignto
\bgroup \tabskip\z@ \@arstrut \@preamble \tabskip\z@ \cr}%
\let\@startpbox\@@startpbox \let\@endpbox\@@endpbox
  \if #1t\vtop \else \if#1b\vbox \else \vcenter \fi\fi
  \bgroup \let\par\relax
  \let\@sharp##\let\protect\relax
  \@arrayskip\@preamble}
\def\eqnarray{\stepcounter{equation}%
              \let\@currentlabel=\theequation
              \global\@eqnswtrue
              \global\@eqcnt\z@
              \tabskip\@centering
              \let\\=\@eqncr
              $$%
 \halign to \displaywidth\bgroup
    \eqnumphantom\@eqnsel\hskip\@centering
    $\displaystyle \tabskip\z@ {##}$%
    \global\@eqcnt\@ne \hskip 2\arraycolsep
         $\displaystyle\arraymode{##}$\hfil
    \global\@eqcnt\tw@ \hskip 2\arraycolsep
         $\displaystyle\tabskip\z@{##}$\hfil
         \tabskip\@centering
    &{##}\tabskip\z@\cr}
\def\theequation{\thesection.\arabic{equation}}
\newfont{\hr}{msbm10}
\newfont{\ams}{msam10}

\begin{document}
\begin{titlepage}
\setcounter{footnote}0
\begin{center}
\hfill ITEP/TH-25/97\\
\hfill FIAN/TD-8/97\\
\hfill hep-th/9707120\\
\vspace{0.3in}
{\LARGE\bf Multiscale $N=2$ SUSY field theories,}\\
\vspace{0.1in}
{\LARGE \bf integrable systems and}\\
\vspace{0.1in}
{\LARGE \bf their stringy/brane origin -- I}
\\
\bigskip
\bigskip
\bigskip
{\Large A.Gorsky
\footnote{E-mail address: gorsky@vxitep.itep.ru}$^{,\dag}$,
S.Gukov
\footnote{E-mail address:
gukov@vxitep.itep.ru, gukov@pupgg.priceton.edu}$^{,\ddag,\dag}$,
A.Mironov
\footnote{E-mail address:
mironov@lpi.ac.ru, mironov@itep.ru}$^{,\S,\dag}$}
\\
\bigskip
\begin{flushleft}
$\phantom{gh}^{\dag}${\it ITEP, Bol.Cheremushkinskaya, 25,
Moscow, 117 259, Russia}\hfill\\
$\phantom{gh}^{\ddag}${\it Landau Institute for Theoretical Physics,
Kosygina, 2, Moscow, 117940, Russia}\hfill\\
$\phantom{gh}^{\S}${\it Theory Department, P.N.Lebedev Physics
Institute, Leninsky prospect, 53, Moscow, 117924, Russia}\hfill
\end{flushleft}
\end{center}
\bigskip \bigskip

\begin{abstract}
We discuss supersymmetric Yang-Mills theories with the multiple scales in the
brane language. The issue concerns $N=2$ SUSY gauge theories with massive
fundamental matter including the UV finite case of $n_{f}=2n_c$, theories
involving products of $SU(n)$ gauge groups with bifundamental matter, and the
systems with several parameters similar to $\Lambda_{QCD}$. We argue that
the proper integrable systems are, accordingly, twisted XXX $SL(2)$ spin
chain, $SL(p)$ magnets and degenerations of the spin Calogero system.
The issue of symmetries underlying integrable systems is addressed.
Relations with the monopole systems are specially discussed.
Brane pictures behind all these integrable structures in the IIB and M theory
are suggested. We argue that degrees of freedom in integrable systems are
related to KK excitations in M theory or D-particles in the IIA string
theory, which substitute the infinite number of instantons in the field
theory. This implies the presence of more BPS states in the low-energy
sector.  \end{abstract}

\end{titlepage}

\newpage
\setcounter{footnote}0
\footnotesize
\section{Introduction}
\setcounter{equation}{0}

During the last years it has become clear that the low energy sector of $N=2$
supersymmetric gauge theories \cite{SW1,SW2} is governed by integrable
systems with the finite number of degrees of freedom so that
physical quantities in the gauge theories (couplings and masses) turn out to
be parameters of these finite-dimensional systems. There have been considered
several examples that include the pure gauge theory \cite{GKMMM} as well as
the theories with adjoint \cite{dw} and fundamental matter \cite{GMMM}. In
particular, in \cite{GMMM}
the $N=2$ super-QCD with the number of flavors $n_{f}<2n$ and
the gauge group $SU(n)$ \cite{GMMM} has been considered
as a first example of the
theory with several scales treated in the integrability framework.
In the present paper, we discuss
more examples of the asymptotically free
theories with the multiple scales.
This condition implies the set of possibilities.
Indeed, one can consider the theory with massive
fundamental matter, the theory with many $\Lambda_{QCD}$ type parameters
instead of the single one, and the system with the gauge group being
product of simple factors \cite{W} so that there is a parameter attached to
each factor. We find integrable counterparts to all these examples.

In fact, there are some (apparently different) models among field theories,
supersymmetric monopoles, string theory compactifications which
possess the same integrable structure that naturally adds to this
list another integrable system counterpart. The whole picture
is drawn in Fig.1.

\begin{figure}[h]
\epsfxsize 400pt
\epsffile{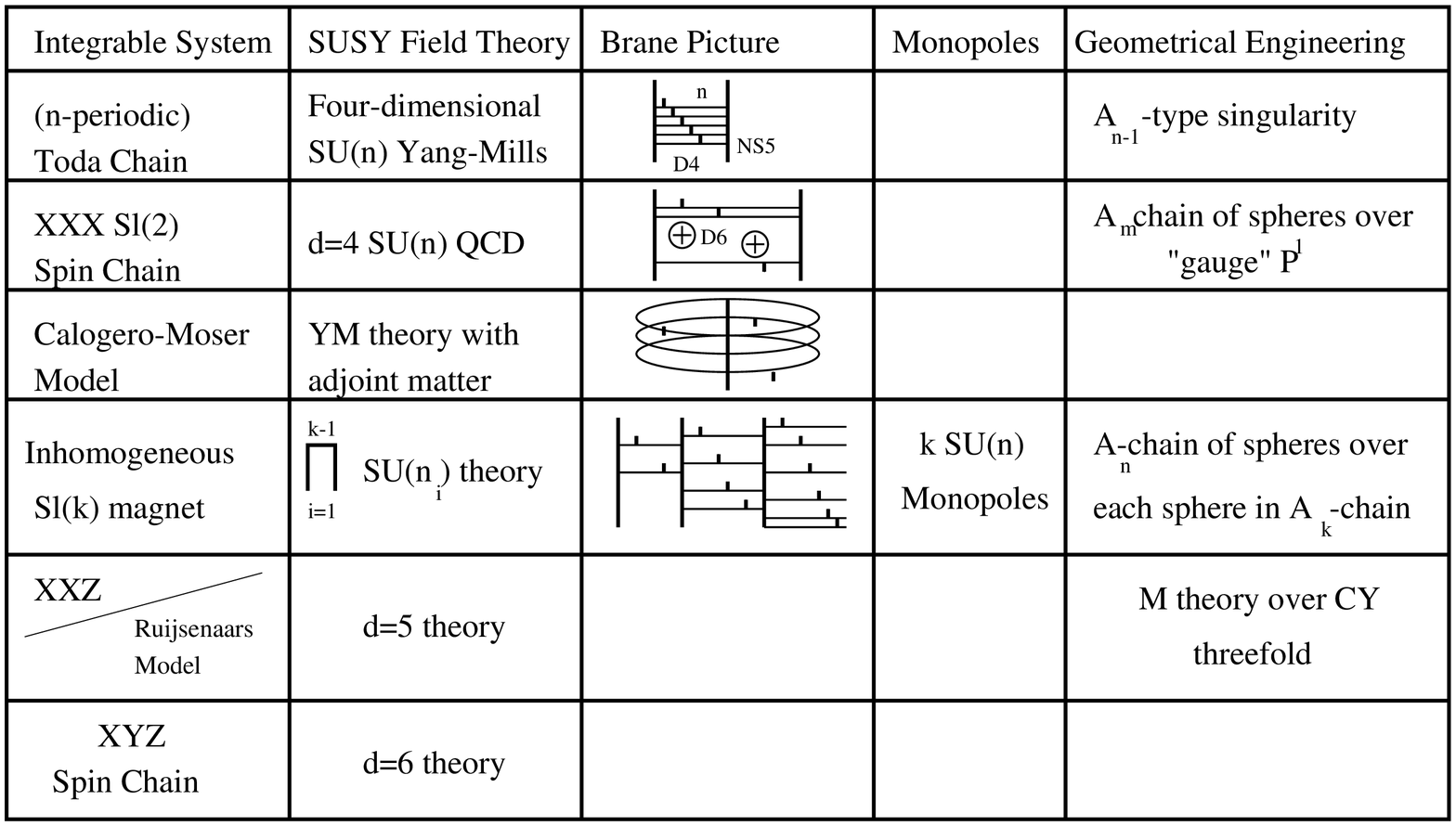}
\caption{Various aspects (columns) of the same models (rows)}
\end{figure}
Let us stress the special role of quiver diagrams in all these
models. The same diagram defines also the corresponding brane
configurations and integrable systems. It also labels the type of singularity
in Calabi-Yau manifold in the geometric engineering approach as well as
S-duality group.

It is clear that the revealing of the hidden integrable structure in the
vacuum sector of SYM theory can not be the aim by itself, thus, we have to
explain the whole motivation of the
approach. First, let us note that the appearance of the integrable system
can be expected from the very beginning. Indeed, integrable system
typically has its configuration or phase spaces realized as some moduli
space. That is what we have in the super-Yang-Mills (SYM) case: in the pure
$N=2$ gauge theory, the Coulomb moduli space and instanton moduli space are
involved. Therefore, we would actually expect a pair of integrable
systems. This is, indeed, the case, and the two integrable
systems -- the Whitham and Hitchin-like ones are exhibited. Identification of
the Whitham dynamics as the one on the Coulomb moduli is quite transparent
but the direct relation of the Hitchin dynamics with the instanton moduli
space is still to be clarified. It has been shown that it is related to the
monopole rather than to instanton dynamics. Still it is easy to connect these
two -- since the monopole solutions can be constructed via an infinite
dimensional counterpart of the instantonic ADHM construction \cite{Nahm}, one
could naturally identify the instantons taken in the conformal t'Hooft anzatz
and organized into $n_{c}$ infinite chains with the circle of $n_{c}$
monopoles \cite{Rossi}. In what follows we use a
kind of Born-Oppenheimer approximation and freeze the ``slow" Whitham dynamics
considering only the ``fast" Hitchin one.

The monopoles which is one of the main subjects of the paper\footnote{Fix
right here that, through the paper, speaking of monopoles, we often imply
just a solution to the Nahm equations without specializing its
boundary conditions. These Nahm equations is the result of applying
the ADHM construction, and, to describe the true monopole, they are to be
added by the proper boundary conditions. We return to the discussion of this
point in s.6.} have a specific nature -- their role is to collect all the
instanton nonperturbative contributions in $4d$ theory.  That is why the
brane configurations we discuss necessarily include the objects which provide
the nonperturbative (instanton) effects in the field theory so that, as a
result, D0-branes within D4-brane gives rise to the nonperturbative
solutions of the monopole type. Note that we add this ingredient to the M
theory approach in comparison with \cite{W}, which corresponds to additional
KK degrees of freedom on the M5 worldvolume.  This is in agreement with the
general description of the finite-gap solutions of integrable systems which
can be formulated as dynamics of points on the spectral curve. Within this
approach the Lax equation is responsible for the structure of the fermionic
zero modes in the monopole background, while the Baker-Akhiezer function
would be associated with the properly normalized fermionic zero mode itself.
The role of the spectral curve is to capture the band structure of the zero
modes which are delocalized in the topologically nontrivial background. Note
that just analogous band structure of the zero modes is expected in the usual
QCD. Let us remark that there are similarities with the scenario of
``confining strings" \cite{Polyakov} which also are the effective objects
which substitute the infinite number of instantons.

Turn now to the role which integrable systems play in the context
of the low-energy SYM theories. First of all, the interpretation
of the field-theory vacuum dynamics in terms of finite number
degrees of freedom immediately raises the question, what these
degrees of freedom really are. Now they are recognized as reflecting the
dynamics of the finite number of interacting branes.
Second, they provide the powerful computational tools to
deal with the vacuum moduli spaces. Although the
D-brane approach looks more transparent, it actually reflects
the degrees of freedom of integrable systems and, therefore, the
quantitative analysis available in the dynamical system seems to be very
useful. For instance,
the phase spaces of integrable systems always have an algebraic
nature, thus, one usually has a huge underlying symmetry group for the
corresponding dynamics \cite{Mir2}. Typically, they are of Yangian or
quantum affine type. It is natural to look for the counterparts of these
symmetries in the brane language. These algebras actually gives a
field theory analog of the string ``BPS like" algebras. We comment
this subject in section 8. Let us remark that, amazingly, this point of
view sometimes can provide even with a more convenient physical insight
\cite{pei}. Another important point about the integrable treatment is that
the finite dimensional integrable systems admit the Lagrangian description
that can be considered as the starting point for the quantization of the
D-brane transverse degrees of freedom.

Let us note that in this paper we intensively use the term ``multiscale" that
we attribute to the three different cases. First, we consider the
$N=2$ theory with fundamental matter where scales are introduced by
the quark masses. The second example deals with the ``group product"
case with the distance between NS5 branes taken as
dimensional parameters\footnote{These parameters can be associated with the
finite cut-off in the UV finite theories or with the dimensional parameters
of $\Lambda_{QCD}$-type in the asymptotically free ones.}. And, finally, the
set of parameters can enter the low energy sector via the multiple
regulator parameters due to a proper dimensional transmutation procedure.

The paper is organized as follows. At first, we review the picture
behind the theory with a single scale and discuss the relevance
of the brane degrees of freedom for integrability. This is done in section
2. In section 3 we complete the identification of the integrable structure
for the $SU(n)$ theory with fundamentals and show that the theory with
$n_{f}=2n$ is governed by the twisted XXX $SL(2)$ spin chain. Section 4 is
devoted to the clarification of the role of higher magnets as the integrable
counterparts for the ``group product" family of theories.
In the
course of our investigation, we works mainly with the ``lattice" Lax
representations \cite{FT} that can be characterized by the quadratic Poisson
brackets with numerical $r$-matrix. Still, to deal with the multiscale
theories, we need ``large" Lax representation of the spin Calogero type that
is characterized by the linear Poisson brackets -- maybe with dynamical
$r$-matrix (this is, of course, nothing but Hitchin system type representation
\cite{Hitsys,nikita2,rub}). Quite astonishing, in many integrable systems one
can meet these both Lax representations so that they lead to the same
spectral curve.  This fact, along with the different Lax representations and
their interpretation in brane terms is contained in section 5.
The fact that
higher magnets emerge not accidentally gets more clear in section 6, where
the correspondence between spin chains and monopoles is established. A new
type of gauge theories with multiple $\Lambda$ type parameters which is
described by degenerations of the spin Calogero (Toda) system is
introduced in Section 7. We also discuss there the corresponding brane
interpretation. Some comments on the hidden symmetry groups are presented in
section 8. In
the Appendix, we demonstrate some very explicit formulas for the $SL(3)$
magnet that are aimed to illustrate some statements of sections 4 and 5
concerning the ``group product" case.

\section{Single scale theory}
\setcounter{equation}{0}

\subsection{Generalities}

We start with the general description of the single scale
SUSY gauge theories.
Hereafter SUSY YM
theories are treated as the theories on the D-brane world-volumes. The
starting point is the consideration of $n$ parallel D-branes embedded in
higher dimensions.  In order to produce 4d theory, this world has 3+1
noncompact dimensions while the coordinates of D-branes in other dimensions
are identified with the vacuum expectation values of the scalar components in
the adjoint hypermultiplets of the theory on the world volume. In fact, some
of these D-brane coordinates in the transverse
directions provide the phase space for the integrable many-body problems.

On the field theory side we consider
the softly broken $N=4$ SYM theory. This theory in the $N=1$ language has
three adjoint hypermultiplets whose complex scalar components parametrize
the remaining six dimensions. The potential energy of these scalars contains
the term $V=\frac{1}{g^{2}}\Tr\sum_{i=1,2,3}[\phi_{i},\phi^{+}_{i}]^{2}$
coming from the kinetic term and the contribution from the superpotential
$W=\Tr\Phi_{1}[\Phi_{2},\Phi_{3}] +M^{2}\Tr(\Phi_{2}^{2}+\Phi_{3}^{2})$,
where $\Phi_{i}$ are the $N=1$ adjoint superfields and $M$ is the mass of
the hypermultiplet. Minimization of the potential energy
provides classical vacuum configurations that, in our case, correspond to the
three scalar fields satisfying the nontrivial commutation relations. Note
that the large mass term in the Lagrangian freezes the fluctuations
transverse to the surface
\be
\Tr(\Phi_{2}^{2}+\Phi_{3}^{2})=const
\ee
which is  an algebraic curve in the two-dimensional complex space.

In this section we will try to fix the degrees of freedom of the
integrable systems in brane terms. Actually, we need to identify
coordinates, momenta and coupling constants in integrable many-body systems.
We will show that the momenta in the integrable system are related
to the positions of D4-branes in the proper direction. These branes
provide the worldvolume for the d=4 $N=2$ theory. Coordinates in the
integrable system come from positions of other branes responsible for the
nonperturbative effects \cite{bsty}, while the coupling constants
correspond to the regulator masses (say, the mass of adjoint scalar in the
Calogero case).

One more ingredient of the integrability to be explained concerns
the meaning of the Lax equation and the Baker-Akhiezer
function. Let us turn, for the sake of simplicity, to the
Toda chain case. There are some
arguments in favor of
interpretation of the Baker-Akhiezer function as the
fermion zero mode in a monopole background. Indeed, the eigenvalue
problem
acquires the form of the equation for the fermionic zero mode ``Hamiltonian"
and similar to the Peierls model treatment
\be
c_{i}\Phi_{i+1}+c_{i-1}\Phi_{i-1}+p_{i}\Phi_{i}=\lambda \Phi_{i}
\ee
where $c_{i}=\exp(q_{i+1}-q_{i})$ and boundary condition
$\Phi_{i+n}(\lambda)=\exp(in\pi(\lambda))\Phi_{i}(\lambda)$ are imposed,
$\pi(\lambda)$ being quasi-momentum.
The elements
of the Lax operator admit the interpretation as matrix elements of the
spectral parameter operator
\be\label{ff}
c_{i}=\int \Phi_{i+1}^{+}\lambda \Phi_{i}d\lambda \\
p_{i}=\int \Phi_{i}^{+}\lambda \Phi_{i} d\lambda
\ee
The spectral curve with this interpretation plays the role of the
dispersion law for the delocalized fermion zero modes.

There are two equivalent ways of constructing Seiberg-Witten
solutions from string theory.
The relation between them was discussed in \cite{V1,ab1,ab2}.
The first one, called geometric
engineering, deals with the Calabi-Yau threefold compactification in
the vicinity of singularity \cite{eng,V1}. The other one is the field theory on
the worldvolume of D4-branes stretched between parallel
NS5-branes \cite{W}:

\begin{equation} \ \ \ \left\{
\begin{array}{c|cccccccccc}
      & 0& 1& 2& 3& 4& 5& 6& 7& 8& 9\cr
NS5   & +& +& +& +& +& +& -& -& -& -\cr
D4    & +& +& +& +& -& -& +& -& -& -\cr
D6    & +& +& +& +& -& -& -& +& +& +
\end{array}\right.
\label{tabl}
\end{equation}
where plus stands for extended direction of each brane and minus
denotes transverse direction.
Since fourbranes are finite in the $x^6$ direction, the long-range
theory is a four-dimensional $N=2$ gauge theory as we need. In the
IIB case, the original supersymmetry (32 real supercharges)
is broken down to $N=2$ in $d=4$ (8 real supercharges) via
geometry of the Calabi-Yau manifold. D-branes, wrapped around shrinking
cycles, correspond to BPS states that further break half of the
supersymmetries. In the IIA case, the SUSY is broken by inserting branes
into the (perturbatively) flat space\footnote{For models without D6-branes.}.
The BPS states in this picture are membranes of M theory ending on a single
NS5-brane. For the sake of unity that we are trying to reach in this
paper, we consider both the Type IIA/M theory and the Type IIB theory in
parallel.

\subsection{Type IIB perspective}

If one deals with the Type IIB/F theory,
$4d$ $SU(n)$ theories without matter emerge as the world volume
theories of the D7-branes wrapped around an elliptically fibered K3.
In fact, the whole picture in this case is not so clear. Still some of the
important ingredients can be observed.
In particular, the key ingredient of the
integrable system -- Calogero model (see, e.g., \cite{Hitsys} and notations
therein) -- namely, the bare spectral curve (torus) whose modulus is the bare
coupling constant enters the picture automatically. Another crucial
ingredient -- holomorphic 1-form $\Phi$ that, after all, turns out to be the
Lax operator corresponds to the D7-branes while a particular
evaluation representation at the marked point is represented by the
background branes. The $n$ D7-branes are wrapped around the bare torus,
therefore, their coordinates in the transverse directions are captured by the
matrix $\Phi$, if the torus is embedded into the two-dimensional complex
surface as the supersymmetric cycle \cite{bsv}. The degrees of freedom
coming from the antiholomorphic form $\bar{A}$ (whose diagonal elements are
the coordinates of the Calogero particles) can be attributed to the branes
representing nonperturbative effects. Finally, the regulators enter the
problem via the insertion of a Wilson line\footnote{This Wilson line
provides the spin variables in the spin Calogero model \cite{nikita2,rub} and
reduces to just a coupling constant in the standard (spinless) Calogero
model.} in the proper representation at the fiber torus in K3. Eigenvalues of
the Wilson line provide the scales in the theory. Proper treatment of the
regulators is still incomplete but some identification in the orientifold
terms \cite{sen} seems attractive.

Let us emphasize that the role of the 11-12 dimensions in F theory in this
consideration is only to provide the necessary input of the symmetry structure,
that is, affine algebra and its representation as well as
the bare coupling constant of d=4 theory. Note also that, in this
picture, the branes are wrapped around the bare not the actual spectral curve.

To provide some additional feeling about degrees of freedom of the integrable
system let us consider the almost trivial ``free" many-body integrable
system. It corresponds to the
d=5 theory without adjoint matter \cite{Nek}. The convenient
look at this system follows from the Chern-Simons
(CS) interpretation of the Ruijsenaars
system, now without interaction. The proper picture \cite{G/G} can be presented
in terms of the
$SU(n)$ bundle over torus if one considers the CS theory on the product
of this torus and an interval. The relation for monodromies around the cycles
$g_{A},g_{B}$ reads
$$
g_{A}g_{B}g_{A}^{-1}g_{B}^{-1}=1
$$
and the diagonal elements of matrices $g_{A},g_{B}$ which, in the free case,
can be diagonalized simultaneously represent the coordinates and the momenta
of the integrable system $g_{A}=diag(e^{ix_{i}});g_{B}=diag(e^{ip_{i}})$.
Remark that, from the F theory point of view, we have degenerated the fiber
torus to a cylinder.

Let us compare this data with the brane picture. There
are two coordinates in the brane picture relevant for the d=5 theory
(let us denote them $x^{5},x^{6}$),
the typical size of the brane configuration in
the $x^{5}$ direction being $R_{5}^{-1}$ and that in the $x^{6}$ direction --
$\frac{1}{g^{2}}$ \cite{ah}. We assume that
the both coordinates are compactified
onto
the circles with radii above. Now our claim is that the tori in the
Ruijsenaars model and in branes are related by the T duality transformation
in the both directions.
Therefore, the radii in the Ruijsenaars torus are
$\alpha^{'}R_{5} ,\alpha^{'}g^{2}$. According to the general rules, the
eigenvalues of the monodromies around the cycles are transformed into
the positions of $n$ branes along the corresponding directions. To be
precise, now we have $n$ 5D branes of the IIB theory with
the (worldvolume) $x^{6}$-positions
which are localized at $p_{i}$ in the $x^{5}$ direction
and $n$ D1 branes with $x^{5}$ worldvolume
coordinates which are localized at $x_{i}$ along $x^{6}$. It is assumed that
our 4+1 theory is defined on first five coordinates of the worldvolume
$(x^{0},x^{1},x^{2},x^{3},x^{4},x^{6})$ of $n$ parallel D5 branes
and positions of branes $p_{i}$ in the $x^{5}$ direction correspond to the
expectation value of one adjoint real scalar field on its worldvolume.
Let us also note that the arising
configuration is the simplest example of the ``polymer picture" from \cite{ah}
in the case when there are no ``background" branes.

Let us now switch on the interaction in the many-body problem. This means that
we add the massive adjoint hypermultiplet in the field theory approach,
add the marked point
with
a nontrivial monodromy around in
the CS picture, or add ``background" branes
in the string language.
In the CS theory, the monodromy relation modifies to
$$
g_{A}g_{B}g_{A}^{-1}g_{B}^{-1}=g_{c}
$$
where
the monodromy $g_{c}$ corresponds to the $CP(n-1)$ coadjoint orbit at the
marked point. After the T duality transformation, one gets the brane picture
with a nontrivial distribution of the flux of the magnetic field on the
torus surface. Actually, the flux distribution is dictated by the structure
of the monodromy $g_{c}$, and one has equal fluxes M on all plaquettes
except the ones on the ``diagonal" $\delta_{ij}$.
This implies a constant magnetic field
through the torus and $n$ ``vortexes"  with the line structure on the
diagonal which locally compensate the external magnetic field. Due to  the
external field the ``brane polymer"  is deformed and the local deformation on
the i-th site $\delta{p_{i}}(x_{i+1}-x_{i})$ is of order M in agreement with
the Lax matrix structure. Note that the set of fluxes through the torus can
be interpreted as the existence of background branes which are localized at
a single point on a T dual torus.

\subsection{Type IIA/M theory perspective}

One of the possible deals with the Seiberg-Witten theory
is to look for its origin in the M theory branes \cite{W}. In order to get
a four-dimensional gauge theory with eight real supercharges, i.e. $N=2$ SUSY,
we consider the field theory on the world volume of the M5-brane.
Its worldvolume splits into the flat $R^{4}$ which is the space-time of
the four-dimensional theory and a compact surface $\Sigma$ of fixed genus $g$
holomorphically embedded into the four-dimensional part $Q_4$ of the
transverse space.

After compactification onto the M theory circle this single fivebrane looks
as a set of Dirichlet fourbranes stretched between NS5-branes.

Let us briefly remind the basic ideas of the geometrical approach
to the
Seiberg-Witten theories \cite{V1,eng}. In order to get $N=2$ four-dimensional
$SU(n)$ gauge theory, one has to consider the Type IIA string
compactification on the Calabi-Yau threefold near the $A_{n-1}$ singularity.
Blowing up the singularity results in changing the complex
structure of the mirror IIB model:
\be
xy=P_n (w)
\label{e1}
\ee
If we have more than one singularity -- that is, more than one gauge group,
the appropriate picture involves an $A_p$-singularity on the base.
Gradually blowing it up, we reduce its rank by one:
\be
w^p + P_{n_1} (\la)w^{p-1} + xy=0
\ee
At the very end, one gets the ``product of gauge groups":
\be
w^p + P_{n_1} (\la)w^{p-1} + P_{n_2} (\la)w^{p-2} \cdots + xy=0
\ee
Incorporation of the fundamental matter means taking coupling
of the $SU(n_f)$ group to zero, i.e. the size of the two-cycle to infinity.
To have fundamental matter for each of the $SU(n)$ factors, replace
the base $P^1$ by the chain of trivalent vertices, with fibers being
the chains $A_{n_i -1}$ of ``gauge" $P^1$'s and the chain $A_{m_i-1}$ of
matter $P^1$'s over each ``gauge" sphere.

It coincides with the Type IIA picture of Witten's
construction \cite{W} described above.
To ensure they are the same theories,
let us briefly revise the appearance of a fivebrane in geometrical
approach.
According to
\cite{OV},
the $A_{n-1}$-singularity on the fiber over $P^1$ in the Type
IIB string theory is T-dual
to the $n$ coinciding fivebranes in the Type IIA theory. We can regard our
ADE-type singularity as a singularity in the local model of $K3$.
Then (\ref{e1})
in each root of $P_n(w)$ describes a degenerated torus. The monodromy
$\tau \rightarrow \tau + 1$ around such singular fiber corresponds
to $\rho \rightarrow \rho + 1$, i.e. $B \rightarrow B + 2 \pi$ from
the dual IIA point of view \cite{OV,vafa1}. In order to see that
this region carries the unit $H$ charge, integrate it over a three-cycle
consisting of a circle $S^1$ around the point with degenerated fiber times
the fiber $T^2$:
\be
{1 \over 2 \pi} \int_{T^2 \times S^1} H =
\int_{T^2 \times S^1} {dB \over 2 \pi}  =
{\Delta B \over 2 \pi} \mid_{S^1} = 1
\ee

It has a straightforward generalization useful for the group product case
that each singular point on the base corresponds to a $NS5$-brane in the
brane language \cite{V1}. These singularities (fivebranes) have also
clear interpretation in integrable models. Their counterparts are
just marked points on the bare spectral curve. Although the elliptic models
are not the subject of the present paper, they give an insight of this
connection. Consider the class of Hitchin systems defined on
torus as the bare spectral curve: (spin) Calogero model, (spin) Gaudin
system, etc. Our conjecture is that all these models have a corresponding
brane interpretation with marked points corresponding to the $NS5$-brane.

Another picture is
actually formulated by Witten \cite{W} (see also \cite{vafa1})
who suggested to treat all branes in the IIA configuration as projections of a
single
M5 brane wrapped around the full spectral curve (not bare one as in the IIB
picture) via different sections.

Let us remark the only notion of the spectral curve is not sufficient
to talk of integrability.
Actually, we need an integrable dynamics that can be realized on
a set of points ($n$ for $GL(n)$ theory) moving on the spectral curve.
In fact, it is right their dynamics that is integrable and is linearized on
the Jacobian and just these degrees of freedom have been missed so far.

Thus, in order to
get a complete integrable system we need the new ingredient
-- a set of points on the
spectral curve which can be represented by the additional KK excitations.
In $10d$ IIA picture they correspond to the nonperturbative corrections
which are interpreted as D0 branes on D4 branes.  But of course it is not a
whole story -- one has to introduce $\Lambda_{QCD}$ via a set of ``magnetic
fluxes". The integrable system we are trying to get in the brane terms and
which governs the answer in the $N=2$ theory without matter is the
periodic Toda chain. It is convenient to start with the theory
with adjoint matter and send the mass of matter to infinity. Then, due to the
dimensional transmutation phenomena a new scale --
$\Lambda_{QCD}=M\exp(-\frac{n}{g^{2}(M)})$ emerges.

In the IIA brane picture, we have $n$ D4 branes stretched between two NS5
branes. These D4 branes provide the worldvolume for d=4 theory.
The $x^{4}+ix^{5}$-coordinates of D4 perturbatively provide
the momenta in the integrable system. There are
also $n$ D0 branes on D4 branes which would be ordered along the $x^{6}$
direction. In the Toda case, their coordinates can be naturally identified
with
\be
x_{j}=j\Delta +\phi_{j}
\ee
where $\Delta$  is a constant which defines the equilibrium position,
$\phi_{i}$ represent the small fluctuations corresponding to the Toda
coordinates and the variables $x_{i}$ can be considered as the coordinates
in the Calogero system (which has no any ``lattice structure", that is to
say, ordering). Note also that, by comparing to the field theory data, one
gets $n\Delta=\frac{1}{g^{2}(M)}$ making our interpretation
along $x^{6}$ direction natural.
D0 on D4 behaves like a monopole,
so we obtain something like the ``monopole ring"
which lies on the ``diagonal".

The possible explanation of the local Toda Lax operator now looks as follows.
The element $L_{11}$ tells that the D0 is placed on the D4
brane so we can assign the same $x^{4}+ix^{5}$ coordinate to it.
Actually the whole issue of the Lax operators
concerns the structure of the  fermionic zero modes in the
topologically nontrivial background, so the very possibility of
$2\times 2$ representation tells that we have no more then 2 fermionic
zero modes here.
The elements $L_{12},L_{21}$ correspond
to the nontrivial matrix elements of the spectral
parameter (\ref{ff}) between the fermionic zero modes localized on the
neighbor D0 branes. At last, we would conjecture that vanishing of $L_{22}$
reflects the fact that only one
chiral zero mode exists for a monopole under consideration.

In what follows we exploit this monopole-related interpretation
of the brane picture. Let us emphasize that we would fairly talk about the
monopoles in the $SU(n)$ theory not about $n$ monopoles of the $SU(2)$ one,
the latter seems to appear because of some duality. Later we will confirm that
just these $SU(n)$ monopole system can be immediately generalized in a more
complicated situation. Let us also remark that evidently we have no the
$2\times 2$ monopole picture for the Calogero system which is in a perfect
agreement with the absence of the second Lax representation (see section 5).

To complete the section outline once again that in any picture
three main ingredients have to be included.
First, one guesses the set of branes
which have to provide the ultimate
worldvolume for the d=4 theory with $SU(n)$ gauge group
-- these are $n$
D7 in the IIB/F or M5 in the IIA/M pictures. Second, one has to add
``instantons" which are M2 and KK excitations in the M theory and finally the
regulator degrees of freedom are introduced -- these are Wilson line on the
torus in IIB/F or some ``magnetic flux" analogues in M theory.

\section{Simple gauge group and $SL(2)$ spin chains \label{magnet}}
\setcounter{equation}{0}

In the section we discuss integrable systems that correspond to some $N=2$
SUSY YM theories. Among these theories there are a few which
are UV-finite. Integrable systems corresponding
to the  UV finite theories softly perturbed by mass terms serve
as ``reference systems" for us, since all other
(asymptotically free) theories can be obtained from the UV-finite ones by
some degenerations.

The defining property of integrable model behind UV-finite theory is that
it is associated with elliptic ``bare curve" whose modulus $\tau$ plays the
role of the bare coupling constant
in the theory. In some cases (theory with adjoint matter
hypermultiplet \cite{dw}) this bare curve
is an underlying Riemann surface where the
Lax operator is defined. In other cases,  considered in
this paper, some coefficients (coupling constants) turn to be some modular
forms of $\tau$. In this latter case, one can ignore the bare curve in the
four-dimensional consideration, just keeping in mind that it lives in the
11-12 dimensions of F theory and dealing
with the coupling constants as just numbers (that are properly adjusted when
doing the dimensional transmutation).

We start with the simplest case of the theory with the $SU(n)$ gauge
group. The integrable structures corresponding to these case are well-known
\cite{GKMMM,dw,GMMM} and we reproduce it here merely
in order to have some basic simple example.

This theory has zero $\beta$-function if one adds $2n$
massive matter hypermultiplets in
the first fundamental representation. Before going into details let us note
that in the brane picture \cite{W} this case corresponds to the two parallel
NS5-branes with $n$ D4-branes stretched between them and $2n$ additional
D6-branes placed between NS5 that provides the UV-finiteness of the
theory. In accordance with our general rules, it corresponds to the
``reference" integrable system. This integrable system is the periodic
inhomogeneous $SL(2)$ XXX spin chain with period $n$ \cite{GMMM} has to
be slightly twisted to make possible the dimensional transmutation to the
asymptotically free theories. Let us note that, although throughout the
paper we use the term ``inhomogeneous $SL(p)$ magnet", it is equivalent, for
the rational (XXX) case, to the homogeneous $GL(p)$ magnet.

\subsection{UV-finite theory}
The periodic inhomogeneous $sl(2)$ $XXX$ chain of length $n$ is given by the
$2\times 2$ Lax matrices
\be\label{LaxXXX}
L_i(\lambda) = (\lambda+\lambda_i)
\cdot {\bf 1} + \sum_{a=1}^3 S_{a,i}\cdot\sigma^a
\ee
$\sigma^a$ being the standard Pauli matrices and $\lambda_i$ being
the chain inhomogeneities, and periodic boundary conditions.
The linear problem in the spin
chain has the following form
\be\label{lproblem}
L_i(\lambda)\Psi_i(\lambda)=\Psi_{i+1}(\lambda)
\ee
where $\Psi_i(\lambda)$ is the two-component Baker-Akhiezer function.
The periodic boundary conditions are easily formulated in terms
of the Baker-Akhiezer function and read as
\be\label{pbc}
\Psi_{i+n}(\lambda)=-w\Psi_{i}(\lambda)
\ee
where $w$ is a free parameter (diagonal matrix).
One can introduce the transfer matrix shifting $i$ to $i+n$
\be\label{Tmat}
T(\lambda)\equiv L_n(\lambda)\ldots L_1(\lambda)
\ee
which provides the spectral curve equation

\be\label{specurv}
\det (T(\lambda)+w\cdot {\bf 1})=0
\ee
and generates a complete set of
integrals of motion.

Integrability of the spin chain follows from
{\it quadratic} r-matrix relations (see, e.g. \cite{FT})
\be\label{quadr-r}
\left\{L_i(\lambda)\stackrel{\otimes}{,}L_j(\lambda')\right\} =
\delta_{ij}
\left[ r(\lambda-\lambda'),\ L_i(\lambda)\otimes L_i(\lambda')\right]
\ee
with the rational $r$-matrix
\be\label{rat-r}
r(\lambda) = \frac{1}{\lambda}\sum_{a=1}^3 \sigma^a\otimes \sigma^a
\ee

The crucial property of this relation is that it
is multiplicative and any product like (\ref{Tmat})
satisfies the same relation
\be\label{Tbr}
\left\{T(\lambda)\stackrel{\otimes}{,}T(\lambda')\right\} =
\left[ r(\lambda-\lambda'),\
T(\lambda)\otimes T(\lambda')\right]
\ee

The Poisson brackets of the dynamical variables $S_a$, $a=1,2,3$
(taking values in the algebra of functions)
are implied by (\ref{quadr-r}) and are just
\be\label{Scomrel}
\{S_a,S_b\} = -i\epsilon_{abc} S_c
\ee
i.e. $\{S_a\}$ plays the role of angular momentum (``classical spin'')
giving the name ``spin-chains'' to the whole class of systems.
Algebra (\ref{Scomrel}) has an obvious Casimir function
(an invariant, which Poisson commutes with all the spins $S_a$),
\be\label{Cas}
K^2 = {\bf S}^2 = \sum_{a=1}^3 S_aS_a
\ee

The spectral curve (\ref{specurv}) can be presented
in more explicit terms:
\be\label{scsl2m}
w^2+\Tr T(\lambda) w+\det T(\lambda)=0
\ee
In accordance with (\ref{Wcurve}), the last term defines masses of the
hypermultiplets. Since
\be\label{detTxxx}
\det_{2\times 2} L_i(\lambda) = (\lambda+\lambda_i)^2 - K^2
\ee
one gets
\be
\det_{2\times 2} T(\lambda) = \prod_{i=1}^n
\det_{2\times 2} L_i(\lambda) =
\prod_{i=1}^n \left((\lambda + \lambda _i)^2 - K_i^2\right) = \\
= \prod_{i=1}^n(\lambda - m_i^+)(\lambda - m_i^-)
\ee
where we assumed that the values of spin $K$ can be different at
different sites of the chain, and
\be
m_i^{\pm} = -\lambda_i \pm K_i.
\label{mpm}
\ee

While the determinant of monodromy matrix (\ref{detTxxx})
depends on dynamical variables
only through Casimirs $K_i$ of the Poisson algebra, the dependence of
the trace $\Tr_{2\times 2}T(\lambda)$ is less trivial.
Still, it depends
on $S_a^{(i)}$ only through Hamiltonians of the spin chain (which are not
Casimirs but Poisson-commute with {\it each other}) -- see further details in
\cite{GMMM}.

Let us note that we have some additional freedom in the definition of the
spin chain and the spectral curve. Namely, note that $r$-matrix (\ref{rat-r})
is proportional to the permutation operator $P(X\otimes Y)=Y\otimes X$.
Therefore, it commutes with any matrix of the form $U\otimes U$. Thus, one
can multiply Lax operator of the spin chain by arbitrary constant matrix
without changing the commutation relations and conservation laws.
Moreover,
we can also insert a constant (external magnetic field) matrix $V$ into the
end of the chain (into the $n$-th site). This is the same as to consider more
general boundary conditions -- those with arbitrary matrix $V^{-1}$. This is
why such a model is called twisted.

The described freedom allows one to fit easily the form of the spectral curve
proposed in \cite{SW2,AS}
\be\label{AS}
w-{Q(\lambda)\over w}=P(\lambda),\\ P(\lambda)=\prod_{i=1}^n
(\lambda-\phi_i),\ \ \ \ Q(\lambda)=h(h+1)\prod_{j=1}^{2n}
\left(\lambda-m_j-{2h\over n}\sum_i m_i\right),\\
h(\tau)={\theta_2^4\over \theta_4^4-\theta _2^4}
\ee
where $\tau$ is the bare curve modular parameter and $\theta_i$ are the
theta-constants \cite{BE}.

It can be done, e.g., by
choosing matrices $U$ and $V$ to be\footnote{To fit (\ref{AS}),
we also need to shift $\lambda_i\to\lambda_i-{2h\over n}\sum_i m_i$.}
\be\label{U}
U_i=\left(
\begin{array}{cc}
1&0\\
0&\alpha_i
\end{array}
\right)\ ,\ \ \ \ \
V=\left(
\begin{array}{cc}
1&h(h+1)\\
\prod_i\alpha_i&0
\end{array}
\right)
\ee
i.e.
\be
V^{-1}={1\over\det V}\left(
\begin{array}{cc}
0&h(h+1)\\
\prod_i\alpha_i&-1
\end{array}
\right)
\ee

Let us remark that further extension of the theories with
the fundamental matter would lead to the (twisted) XXZ and XYZ chains for
the 5d and 6d theories correspondingly.

\subsection{Pure gauge theory\label{Toda}}
Now let us briefly consider the degeneration of our reference system to the
$n_{f}<2n$ case.
This can be done in the
standard way \cite{SW2} by taking $l$ masses $m_1,...,m_l$ to infinity
while keeping $\Lambda_{QCD}^l\equiv e^{i\pi\tau}m_1...m_l$ finite.
After this procedure the modular forms disappear so that
$\Lambda_{QCD}$ emerges instead.

Degenerations of the system can be studied
at a single site (for the sake of brevity, we omit the index of
the site). Let us consider the Lax operator $\tilde L=UL$ (see
(\ref{LaxXXX}), (\ref{U})) with spins satisfying the Poisson brackets
(\ref{Scomrel}). We are going to send $\alpha$ to zero. Along this line
of reasoning, there are
two possibilities to get nontrivial Lax operator. The first possibility,
when the both masses (\ref{mpm})
disappear and one reaches the pure gauge theory, is described by
the periodic Toda chain \cite{GKMMM}. In order to get it, one needs to
redefine $S_+\to {1\over\alpha}S_+$, then take
$\alpha$ to zero and remove, after this, the inhomogeneity by the shift of
$S_0$. This brings us to the Lax operator of the form
(we introduce new notations $S_0=S_3$, $S_{\pm}=S_1\pm iS_2$)
\be
\left(
\begin{array}{cc}
\lambda+S_0&S_-\\
S_+&0
\end{array}
\right)
\ee
so that the Poisson brackets are
\be
\left\{S_{\pm},S_0\right\}=\pm S_{\pm},\ \ \ \ \left\{S_+,S_-\right\}=0
\ee
This algebra is realized in new (Heisenberg) variables $p$ and $q$
\be
S_{\pm}=\pm e^{\pm q},\ \ \ \ S_0=p,\ \ \ \ \{p,q\}=1
\ee
This leads us finally to the Toda chain Lax operator \cite{FT}
\be
L_{Toda}=\left(
\begin{array}{cc}
\lambda+p&e^q\\
-e^{-q}&0
\end{array}
\right)
\ee
and the spectral curve (\ref{scsl2m}) takes the form
\be\label{scToda}
w^2+\Tr T(\lambda) w +1=0
\ee

Now let us return to the second possibility of the asymmetric degeneration,
when one of the masses (\ref{mpm}) remains in the spectrum
while the second one goes to infinity. One can understand from (\ref{mpm})
that, in contrast to the Toda case, this degeneration requires a special fine
tuning of the Casimir function and inhomogeneity, so that both of them go to
infinity but their sum (difference) remains finite. In the Lax operator terms
it can be done in the following way. Let us redefine $S_+\to {1\over\alpha}S_+$
{\it and} $S_0\to {1\over\alpha}S_0$. This means that the Poisson brackets
take the form
\be\label{specal}
\{S_+,S_-\}=-2S_0, \ \ \ \ \{S_{\pm},S_0\}=0
\ee
Now in order to preserve finite Lax operator (\ref{LaxXXX}), one needs to
take care of its element $L_{11}(\lambda)$. This can be done by the rescaling
$\lambda_i\to {1\over\alpha}\lambda_i$ and fixing
$\lambda_i+S_0$ to be $\alpha\cdot s_0$. This brings us to the
Lax operator
\be\label{Laxi}
L(\lambda)=\left(
\begin{array}{cc}
\lambda+ s_0& S_-\\
S_+& \lambda_i-S_0
\end{array}
\right)
=\left(
\begin{array}{cc}
\lambda+s_0& S_-\\
S_+& -2S_0
\end{array}
\right)
\ee
The determinant of this Lax operator is equal to $(\lambda -m)$ where $m$
is the finite mass
\be
m=2s_0S_0+S_+S_-
\ee
in perfect agreement with (\ref{mpm}),
Let us note that $2s_0S_0+S_+S_-$ is also the Casimir
function of the algebra (\ref{specal}), since $\{S_{\pm},s_0\}=\pm S_{\pm}$.

In fact, the Lax operator (\ref{Laxi}) can be rewritten in the form
\be
L(\lambda)=\left(
\begin{array}{cc}
\lambda+ s_+s_--m& s_-\\
s_+& 1
\end{array}
\right)
\ee
where we rescaled the spins and used the explicit form of fixed
the Casimir function $2s_0S_0+S_+S_-$. This Lax operator defines so-called
novel hierarchy \cite{NI} and is gauge equivalent to the discrete AKNS
\cite{DAKNS} and  relativistic Toda chain \cite{KMZ}. This latter
correspondence looks especially unexpected, since the same relativistic Toda
chain describes the 5d pure gauge SUSY theories \cite{Nek}.

Let us note that another (equivalent) way to count all possible degenerations
\cite{Khar} is to consider $2\times 2$ Lax operator of the most general
form linear in $\lambda$, which satisfies the Poisson brackets
(\ref{quadr-r}) with the rational $r$-matrix (\ref{rat-r}) and  determine
all  Casimir functions with respect to this Poisson brackets. Then,
all possible degenerations are in one to one correspondence with the
vanishing of Casimirs\footnote{We are grateful to S.Kharchev for the useful
discussions on this point.}.

To complete this section, let us say a few words of the brane interpretation
of the Toda variables $p_i$. First, restore the dependence on $\Lambda_{QCD}$
in the Toda Lax operator. It can be done in many different ways, we choose it
to be
\be
L_{Toda}=\left(
\begin{array}{cc}
\lambda+p&\Lambda^2_{QCD}e^q\\
-e^{-q}&0
\end{array}
\right)
\ee
thus the spectral curve is of the form
\be
w^2+\Tr T(\lambda)+\Lambda_{QCD}^{2n}=0
\ee
Now taking the perturbative limit, i.e. putting $\Lambda_{QCD}\to 0$, we get
that the spectral curve degenerates to the two sphere touching each other in
the $n$ points that are zeroes of the polynomial $\Tr T(\lambda)$. These
zeroes are equal to $p_i$ in the perturbative limit and describe the
positions of the D4-brane ends in the $(x^4+ix^5)$-plane, in accordance
with the picture of \cite{W}.

There is another class of the spin chain variables, namely, those describing
masses of the sixbranes (\ref{mpm}), which can also allows very rough but
transparent brane interpretation. Indeed, in accordance with (\ref{mpm}), one
can associate with each D4-brane (i.e. site of the spin chain) a pair of the
sixbranes, the distance between them being measured by the value of the
Casimir function $K$. At the same time, the $(x^4+ix^5)$-coordinate of the
center of masses of these two sixbranes is naturally measured by the
inhomogeneity $\lambda_i$. This looks especially reasonable, since just the
inhomogeneity corresponds to the decoupling $U(1)$ factor of the $GL(2)$
magnet.\footnote{One more interpretation of (\ref{mpm}) might look as follows.
We can start with a free theory and would expect the monopoles in ``the
momentum space" by the following reason. For the free particle, there is the
standard dispersion relation $E=\pm ({\vec{p}}^{2} +m^{2})^{\frac{1}{2}}$
which tells us that there is the level crossing on the multi-dimensional
surface $\vec{p}=\pm im$ in the complex momenta space. Would be this surface
one-dimensional, according to the general philosophy of the Berry phase,
this means that one needs to introduce two monopoles in ``the momentum
space"  with their centers located at points $p=\pm im$. This
resembles formula (\ref{mpm}) without inhomogeneities. Furthermore, one can
switch on the interaction with the Higgs field that results into the shift of
the fermion mass due to the Yukawa coupling. Since inhomogeneities are
naturally related to the Higgs vev's, one finally arrives at formula
(\ref{mpm}).}

Note that, within this interpretation, the procedure of the degeneration gets
its pictorial explanation. It just corresponds to removing one of the
sixbranes to infinity. To keep the other branes at finite distances, one
needs to tune parameters $\lambda_i$ and $K$ so that one of the two
quantities $-\lambda_i\pm K$ remains finite.

\section{The group product case}
\setcounter{equation}{0}
\subsection{Reference system}
Consider now more complicated case of the gauge group which is the
product of some $SU(n)$ factors: $G=\prod_{i=1}^{p-1} SU(n_i)$.
One can also add $p-2$ massive hypermultiplets in the
representations $(n_i,\bar n_{i+1})$ and also by some hypermultiplets in the
first fundamental representations. This case was
first considered by E.Witten in \cite{W} and possesses a new important
feature. To start with, let us note that in this case for many different
choices of $n_i$
the suitable number of matter hypermultiplets
can be added
to make $\beta$-function zero. Actually, one needs to add
\be
d_i=2n_i-n_{i-1}-n_{i+1}, \ \ \ n_0=n_p=0
\ee
hypermultiplets \cite{W}. This always can be done unless some $n_i-n_{i+1}$
are too large. However, among all the integrable systems describing
these numerous UV-finite theories there is a unique one that plays the role of
the reference system, since all others can be obtained
by degenerations. This theory is specified by maximal possible
number of the matter hypermultiplets at given $p$ and $n_1$ and characterized
by the set $\{n_i\}=n,2n,3n,\ldots,(p-1)n$. The corresponding number of
the hypermultiplets is $n_1+n_{p-1}=np$.

Now come to the brane picture. In this case we have $p$ parallel NS5-branes
with $n_1$ D4-branes stretched between the first pair of them, $n_2$
D4-branes stretched between the second pair of the NS5-branes etc, with
the corresponding numbers $d_i$ of D6-branes. The reference theory in this
language corresponds to Fig.2.

\begin{figure}
\epsfxsize 300pt
\epsffile{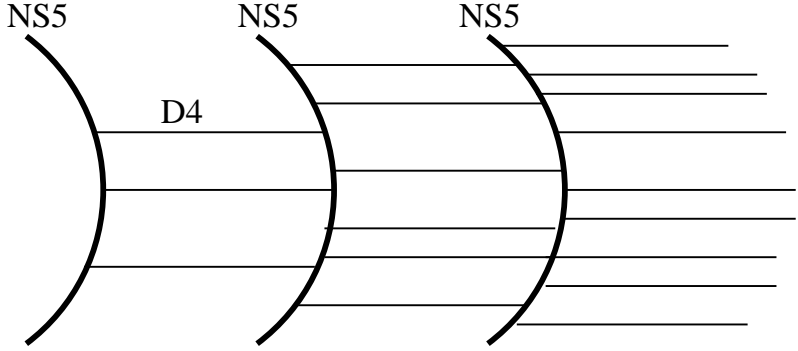}
\caption{Brane picture, corresponding to the $Sl(3)$ magnet}
\end{figure}

Its degeneration to theories with fundamental matter (D6-branes) in
brane languages corresponds that $\la=x^4 + i x^5$ positions
of several fourbranes coincide. Namely, if $k$ of $n_{p-1}$ D4-branes
have the same $\la$ coordinates as $k-1$ of $n_{p-2}$ and so on
till the only D4-brane of $n_{p-k}$ set, we can attach a fictitious
D6-brane to the right infinite end of the D4-set with this value of
$\la=x^4 + i x^5$, and move it through $k-1$ fivebranes to the
left consequently annihilating the whole set of such D4's.
This mechanism will be explained in details in section 5 via brane
creation phenomenon.

In further assigning brane configurations
with integrable systems, it will be of
importance that,
with any described branes system,
there can be associated the spectral curve
of the form \cite{W}
\be\label{Wcurve}
w^p + g_{n_1}(\lambda)w^{p-1}+g_{n_2-d_1}(\lambda)J_1(\lambda)w^{p-2}
+...+\\+ g_{n_k-(k-1)d_1-(k-2)d_2-...-d_{k-1}}(\lambda)
J_1^{k-1}(\lambda)J_2^{k-2}(\lambda)...J_{k-1}(\lambda)w^{p-k}+...+
\prod_k^{p-1} J_k^{p-k}(\lambda)=0
\ee
Here $g_i(\lambda)$ is a polynomial of degree $i$ depending on the
Coulomb moduli\footnote{Hereafter, we call ``Coulomb moduli" the moduli
corresponding to the vev's of the fields in the vector (gauge) multiplet.}

and hypermultiplet masses, and
\be\label{J}
J_i=\prod_{j=d_{i-1}+1}^{d_i}(\lambda-m_j)
\ee
are pure mass terms ($m_j$'s are masses of the hypermultiplets in the
fundamental representations).

Let us note that this spectral curve gives another justification for our
choice of reference system. Indeed, let us consider ``the pure gauge
case"\footnote{To simplify further terminology, hereafter we call the system
that contains no fundamental hypermultiplets ``the pure gauge theory", despite
there are still presented $p-2$ hypermultiplets in the representations
$(n_i,\bar n_{i+1})$. Our terminology refers to the fact of absence of any
D6-branes and semi-infinite branes and takes its origin in the $SU(n)$ case.}
when
all $J_i(\lambda)=1$ \cite{W}.
Then, only the moduli space of the reference
spectral curves contains the $Z_n$-symmetric curve, and, in this sense, is
maximally possible complete (although the moduli space of curves
that correspond to an integrable system is always not complete for
sufficiently large genera and covers typically $g$-dimensional subspace
of the complete $(3g-3)$-dimensional moduli space;
the $g$ moduli are just the Hamiltonians of integrable
system and, therefore, the dimension of the configuration space is
exactly $g$).

To clarify this point, let us introduce the new variable
$Y=w-{g_{n_1}\over p}$. Then,  only with the choice
$n_k=kn$ one can match the polynomials $g_k$ so that the curve takes the form
\be
Y^p=R(\lambda)
\ee
where $R(\lambda)$ is a polynomial of degree $np$, i.e. this curve is of
genus
\be\label{gf}
g=\left({np\over 2}-1\right)(p-1)
\ee
The same genus can be
obtained immediately from the brane picture, if one implies that the
D4-branes are blown up to the handles of the Riemann surface, while
NS5-branes are the spheres that are glued by these handles. Then, $n$
D4-branes between first two NS5-branes give $n-1$ wholes, $2n$ next
branes give $2n-1$ holes etc. -- totally, it is exactly $\left({np\over
2}-1\right)(p-1)$ holes.

In fact, this interpretation of the Riemann surface (spectral curve) as the
brane graph implied in \cite{W} is justified by the
calculation of genus of the general curve
\be\label{Wcurveg}
w^p + g_{n_1}(\lambda)w^{p-1}+g_{n_2}(\lambda)w^{p-2}
+...+ g_{n_k}(\lambda)
w^{p-k}+...+ g_{n_p}(\lambda)=0
\ee
The degrees of polynomials in this expression are restricted by the
conditions $n_1\le n_2\le ...\le n_{p-1}$ and $n_1\ge n_2-n_1\ge n_3-n_2
\ge ... \ge n_p-n_{p-1}$.
The second condition is the requirement for all the $\beta$-functions to be
non-positive, i.e. for the theory to be asymptotically free or UV-finite.
These conditions allow one to calculate easily the genus of the curve. For
doing this, we use the  elegant formula
(see, for example, \cite{Dub})\footnote{We are
grateful to A.Chervov who has drawn our attention to this formula.}
for the genus of the non-singular projective curve given in the
affine map by the polynomial $F(w,\lambda)= \sum a_{ij}w^i\lambda^j$. The
genus is merely equal to the number of the integer points in the Newton
polygon $\Delta$ for $F$, i.e. the convex shell of the points
$(i,j)\in {\bf  Z}$ such that $a_{ij}\ne 0$ -- see Fig.3.\footnote{This
formula is proved by noting that,
for each pair $(i,j)$ inside the polygon, there is the holomorphic
differential \be\label{diff} d\omega_{ij}={w^{i-1}\lambda^{j-1}dw\over
F'_{\lambda}(w,\lambda)} \ee .}

Since powers of $w$ represent resolution of singularity on the base,
and the degree $n_i$ of each polynomial $g_{n_i}$ corresponds to resolution
of $A_{n_i - 1}$ singularity on the fiber, by construction, $\Delta$
is nothing but toric polyhedron responsible for the local geometry
of Calabi-Yau threefold \cite{V1}.

Following \cite{V1} each interior node of $\Delta$ define a compact
divisor, thus generating a homology group.
As it was explained in details in \cite{vafa1}, Riemann surface
$\Sigma$ faithfully represents all the data of Calabi-Yau cycles,
with correspondence between cycles and forms. In other words this
means that each compact divisor counts an element of $H_1(\Sigma)$.

In the present paper, from the integrable
system point of view, and in \cite{V1}, from the geometrical side, it was
independently inferred that to have UV-free four-dimensional theory,
i.e. each $d_i \ge 2n_i - n_{i-1} - n_{i+1}$, means to consider only
convex $\Delta$.

\begin{figure}
\epsfxsize 300pt
\epsffile{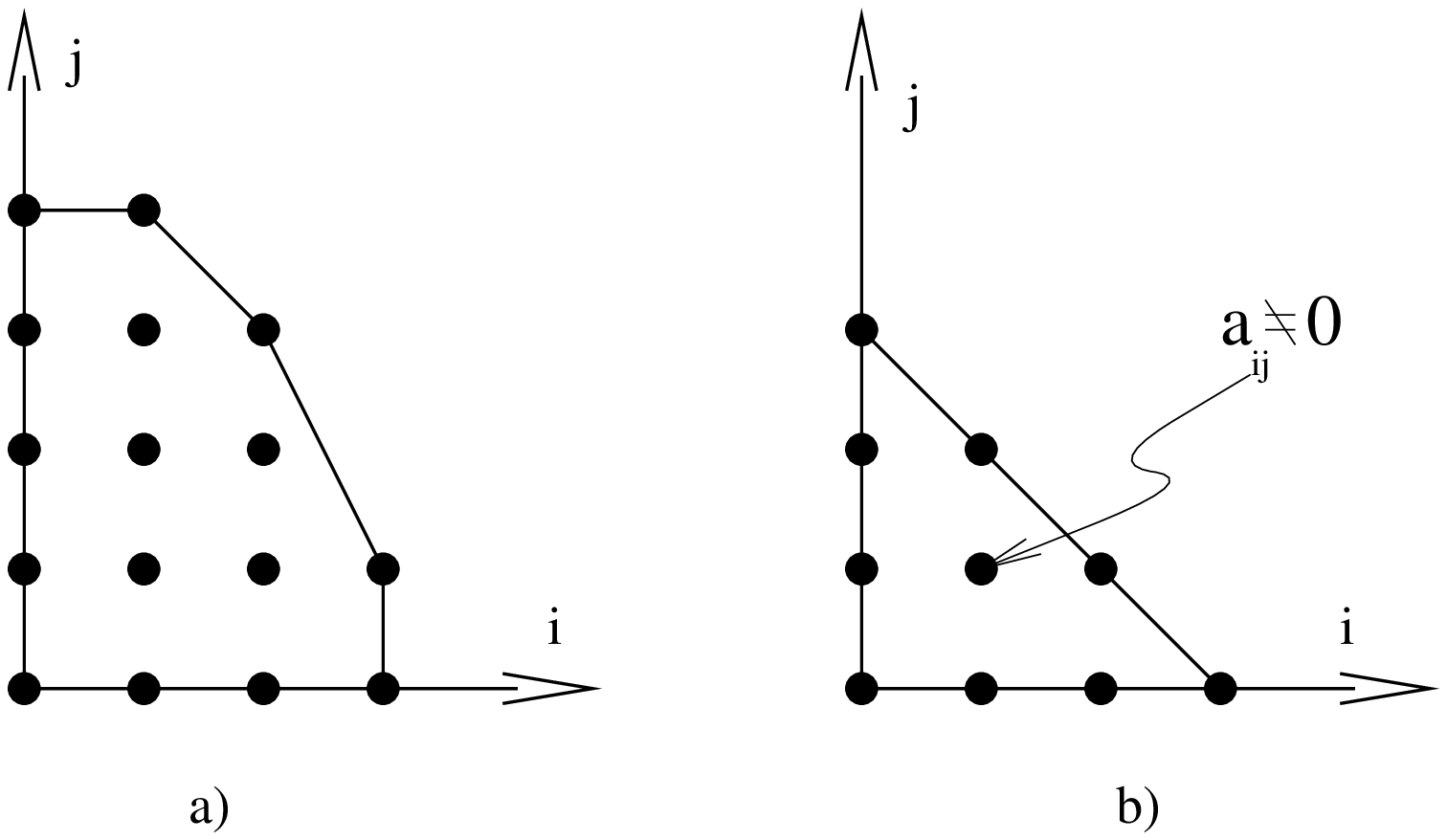}
\caption{Newton polygon $\Delta$}
\end{figure}

Using the conditions onto the set of $\{n_i\}$ (the second condition
just guarantees the convexity), one can easily get that the genus
of the curve (\ref{Wcurveg}) is equal to
\be\label{genus}
g=\sum_{i=1}^{p-1}n_i-p+1
\ee

Let us note that this result coincides with one calculated from the
brane picture as described above, since $n_1$ D4-branes between first
pair of the NS5-branes gives $n_1-1$ holes etc. Remark also that
in accordance with expectations,
the degree of the last term does not
enter the final answer, which means that the genus is not effected by the
D6-branes (matter hypermultiplets) added.

Now, using the formula (\ref{genus}), we can determine the reference system as
corresponding to the curve of the maximal possible genus at given $p$ and
$n_1=n$.\footnote{Note that, by the shift of $w$, $g_{n_1}(\lambda)$ can be
cancelled. For the reference curve, however, this shift does not change the
degrees of all the rest polynomials $g_{n_i}(\lambda)$.} This genus is equal,
in accordance with (\ref{genus}), to $\left({np\over
2}-1\right)\left(p-1\right)$.

Below we describe the reference integrable system that is
the periodic inhomogeneous $SL(p)$ (or homogeneous $GL(p)$) spin chain with
period $n$.  It is described by the curve (\ref{Wcurve}) with all but
$J_{p-1}$ equal to 1.

\subsection{UV-finite theory}
The inhomogeneous $SL(p)$ spin chain is
similar to the $SL(2)$ spin chain considered above
and given by the $p\times p$ Lax operator
\be\label{Lax}
L_i(\lambda)=K^{ab}S_{a,i}X_{a}+(\lambda+\lambda_i)\cdot {\bf 1}
\ee
at the $i$-th site. Here $S_{a,i}$
are dynamical variables\footnote{To reproduce the $SL(2)$ case considered
above, one needs to make the replacement $S_0\to 2S_0$.},
$X_a$ are generators of the $SL(p)$ algebra
and $K^{ab}$ is its Killing form. We always choose below $X_a$ to lie in the
first fundamental
representation of $SL(p)$. In complete analogy with (\ref{lproblem}),
this Lax operator shifts the Baker-Akhiezer function to the neighbor site:
\be
L_i(\lambda)\Psi_i(\lambda)=\Psi_{i+1}(\lambda)
\ee
and periodic boundary conditions read as
\be
\Psi_{i+n}(\lambda)=-w\Psi_{i}(\lambda)
\ee
Introducing the transfer matrix shifting $i$ to $i+n$
\be
T(\lambda)\equiv L_n(\lambda)\ldots L_1(\lambda)
\ee
and using boundary conditions, one gets the spectral curve equation
\be\label{specurvp}
\det (T(\lambda)+w\cdot {\bf 1})=0
\ee
In fact, (\ref{specurvp}) is a polynomial in
$w$ and $\lambda$ with the coefficients that give an ample set of integrals
of motion.

The Poisson structure is in charge of integrability of the system and
is given by (\ref{quadr-r}) with the rational $r$-matrix
\be\label{rg}
r(\lambda)={K^{ab}X_a\otimes X_b\over \lambda}={P\over\lambda}
\ee
where $P$ is the exchange operator $P(X\otimes Y)=Y\otimes X$
and the last equality is correct, since we consider the fundamental
representation. This gives rise to the Poisson brackets of $S_i$
\be
\left\{S_a,S_b\right\}=-f_{ab}^cS_c
\ee
where $f_{ab}^c$ are the structure constants of the $SL(p)$ algebra.
The Poisson structure is, generally speaking, degenerated so that its
annulator is the algebra of Casimir functions that are defined to be
invariant with respect to the coadjoint action.
Therefore, one needs to
restrict the Poisson bracket to the orbits of this action.

Traces of degrees of Lax operators are Hamiltonians which mutually
commute with respect to this Poisson structure. It can be easily understood
since (\ref{quadr-r}) is multiplicative and, therefore, products of Lax
operators satisfy the same Poisson brackets.

Now one can use the explicit formulas for the Lax operator to obtain the
following spectral curve
\be\label{sc}
w^p+g_1(\lambda)w^{p-1}+...+g_k(\lambda)w^{p-k}+...+g_p(\lambda)=0
\ee
where $g_k$ is a polynomial of degree $kn$. Manifestly, $g_1(\lambda)=\Tr
T(\lambda)$, ..., $g_k=\sum_I \det M^{(k)}_I(\lambda)$, ...,
$g_p(\lambda)=\det T(\lambda)$, where $M^{(k)}_I(\lambda)$ is the
matrix obtained from the transfer matrix by removing
$k$ columns and $k$ rows that intersect at the diagonal elements given by the
multi-index $I$.
This spectral curve is exactly (\ref{Wcurve}) with all but $J_{p-1}$
put equal to 1.

Apart from the spectral curve, there is another key ingredient of the
integrable system which is of great importance in the supersymmetric
theories. This is the meromorphic differential $dS$ \cite{GKMMM,GMMM,Mir}
that generates the spectrum of the BPS states in the theory \cite{SW1,SW2}.
The defining property of $dS$ is that its variations over
Coulomb moduli gives
holomorphic 1-forms. In the spin chains, this differential has the form
\cite{WDVV}
\be
dS=\lambda{dw\over w}
\ee
Let us check that this is, indeed, the proper differential, i.e. its
variations over the Coulomb moduli give the holomorphic differentials.
Consider the curve (\ref{sc}) (or (\ref{Wcurveg})) and write it in the form
$F(w,\lambda)=\sum_{i,j=0}a_{ij}w^i\lambda^j=0$.
The differential $dS$ depends
on moduli $a_{ij}$ of the curve only through the dependence of $\lambda$
determined by the equation $F(w,\lambda)=0$. Thus, we get
\be
{\partial dS\over\partial a_{ij}}={\partial \lambda\over\partial a_{ij}}{dw
\over w}={\partial F\over\partial a_{ij}}{dw\over wF'_{\lambda}}=
{w^{i-1}\lambda^j dw\over F'_{\lambda}}
\ee
Now it remains to note that the coefficients $a_{p-k,n_k}$ are
just fixed numbers (unities)\footnote{This property of (\ref{sc}) follows
from the manifest form of the Lax operator (\ref{Lax}).} that are not to be
regarded as moduli, while $g_p(\lambda)=\prod(\lambda-m_i)$ does not contain
the Coulomb moduli and only masses. Then, using (\ref{diff}) we finally get
that the variations of $dS$ over the Coulomb moduli do really generate all the
holomorphic differentials.

In order to check that the inhomogeneous periodic $SL(p)$ spin chain is
really the reference system we are looking for, one needs to check that it
properly degenerates to reproduce the curve (\ref{Wcurve}) with arbitrary
$J_i(\lambda)$. Indeed, from (\ref{sc}) one reads off that $J_{p-1}=\det
T(\lambda)=\prod_i^n\det L_i(\lambda)$. Each determinant $\det L_i(\lambda)$
is a polynomial of degree $p$ with coefficients being Casimir functions. On
the generic orbit,
all zeroes of this polynomial and, therefore, masses are
different. Thus, we obtain $J_{p-1}= \prod_i^n\prod_r^p(\lambda-m_{i,r})$,
where $m_{i,r}$ refers to $p$ different roots of $\det L_i(\lambda)$.

In order to get generic $J_i$, one needs to consider some coinciding
masses {\it at the same site}. If only two masses coincide, one gets only
$J_{p-1}\ne 1$ and $J_{p-2}\ne 1$. This case corresponds to a specific
set of orbits
with one relation between Casimir functions. Since there are $p-1$
($=\hbox{rank } SL(p)$) independent Casimir functions, there can exist
maximally $p-1$ relation between them. The corresponding  special
orbits describe $J_1(\lambda)$. Certainly, at different sites, one can
consider different degenerations, giving $g_p(\lambda)$ the general form as
in (\ref{Wcurve}).

Now this is nontrivial test of the whole construction to check that the
spectral curve (\ref{sc}) with all these degenerations gets the form
(\ref{Wcurve}), i.e. as soon as two roots of the polynomial $g_p(\lambda)$
coincide, the polynomial $g_{p-1}(\lambda)$ gets simple zero at the same point;
as soon as three roots of the polynomial $g_p(\lambda)$
coincide, the polynomial $g_{p-1}(\lambda)$ gets double zero at the same
point, while the polynomial $g_{p-2}$ gets simple zero at this point etc.
This is, indeed, the case and in the Appendix we demonstrate this by
the direct calculation in the simplest case of the $SL(3)$ spin chain.

To conclude this subsection, let us note that the interpretation of the mass
formulas for the sixbranes proposed in sect.3 is naturally continued to the
$SL(p)$ magnet. Now a site of the spin chain should be associated with one
D4-brane suspended between the first pair of the NS5-branes, two D4-branes
suspended between the second pair, ..., $p-1$ D4-branes suspended between the
last pair. With this last set of the $p-1$ D4-branes is then associated $p$
sixbranes, with their coordinates being described by the masses. In this
situation, again, the inhomogeneity describes the $(x^4+ix^5)$-coordinate of
the center of masses of the sixbranes, and bringing one of the branes to
infinity describes the procedure of the degeneration.

\subsection{Pure gauge theory}
Now let us discuss decoupling of the fundamental hypermultiplets, that is
to say, degeneration of the spin chain. As before, the degeneration can be
manifestly done using the freedom of multiplying the Lax operator by any
constant matrix, since the $r$-matrix (\ref{rg}) is again the permutation
operator and commutes with the matrix $U\times U$. This procedure looks
literally the same as in the $SL(2)$ case. Thus, let us now discuss only the
maximally degenerated systems corresponding to the pure gauge theories. These
are described by $SL(p)$-generalizations of the Toda chain system.

As in the $SL(2)$ case, it is sufficient to consider the Lax operator at a
single site. In the $SL(p)$ case there are $p-1$ different possible
degenerations of (\ref{Lax}) generalizing Toda system and $p-1$ different Lax
operators. They contain the spectral parameter $\lambda$ in the only one
diagonal term, in two diagonal terms etc. till $p-1$ diagonal terms and have
unit determinant. Varying this different Lax operators along the chain, one
can reproduce polynomials of different degrees in (\ref{Wcurve}) with all
$J_k(\lambda)=1$ (i.e. all $d_i=0$), the degrees being subject to the
inequalities $n_1\le n_2\le \ldots\le n_{p-1}$ and $n_1\ge n_2-n_1\ge n_3-n_2
\ge ... \ge n_p-n_{p-1}$. This really corresponds to
the pure gauge theory, see \cite{W}.
These different possible polynomials correspond to several possible
partitions of D4-branes between background D6-branes.

As we already showed in s.\ref{Toda}
(very explicit formulas for the $SL(3)$ case
can be found in Appendix), one can
obtain the different degenerations just multiplying the Lax operator
(\ref{Lax}) by constant diagonal matrices $U$ with 1, 2, ..., $p-1$ non-unit
entries equal to $\alpha$, rescaling some spin operators
$S\to{1\over\alpha}S$ and then taking $\alpha$ to zero. Here we demonstrate
the equivalent way of doing briefly mentioned in the very
end of s.\ref{Toda} \cite{Khar}.

Namely, let us start with the general $p\times p$ Lax operator that satisfies
the Poisson brackets (\ref{quadr-r}) with the rational $r$-matrix (\ref{rg}):
\be\label{PB}
\{L_{ij}(\lambda),L_{kl}(\mu)\}={1\over \lambda-\mu}\left(
L_{kj}(\lambda)L_{il}(\mu)-L_{il}(\lambda)L_{kj}(\mu)\right)
\ee
Consider specific
representations that are described by the Lax operators of the form
(we consider, for a moment, the homogeneous $GL(p)$ system instead of
inhomogeneous $SL(p)$ one)
\be\label{LaxY}
L_{ij}(\lambda)=a_i\lambda\cdot{\bf 1}+A_{ij},\ \ \ \ \Tr A=0
\ee
These Lax operators can be produced from (\ref{Lax})
multiplication by a constant matrix. From (\ref{PB}) one can read off the
Poisson brackets
for the matrix $A$
\be\label{cr}
\{A_{ij},A_{kl}\}=a_kA_{il}\delta_{kj}-a_iA_{kj}\delta_{il}
\ee
with $a_i$ being Casimir functions.
From these Poisson brackets one concludes that, for an element $A_{ij}$ to be
a Casimir function, there should be fulfilled the condition $a_i=a_j=0$. Now
let us look at the determinant of the Lax operator (\ref{LaxY}). As  was
explained above it has the form
\be\label{detY}
\det L(\lambda)=a_1a_2...a_p\lambda^p+C_2\lambda^{p-2}+C_3\lambda^{p-3}
+...+C_{p}
\ee
where $C_i$ are the Casimir functions (this formula specifies their choice).
In the general $SL(p)$ magnet, all $a_i\ne 0$ and can be put equal to 1. In
this case, the determinant (\ref{detY}) is a general polynomial of degree $p$
with $p$ zeroes. Degenerating of this system implies that some (say, $k$) of
these zeroes (masses of the hypermultiplets) are taken to infinity, i.e. some
$a_i=0$ and some ($k$) first Casimir functions vanishes too. The pure gauge
theory corresponds to the limiting degeneration when the determinant
(\ref{detY}) is equal to constant (unity). This is reached by putting some
$a_i$ and all Casimir functions but the last equal to zero.
Let us note that there exist exactly as many
as $p-1$ classes, since one can put 1, 2, ..., $p-1$ $a_i$'s equal
to zero. This completes the construction.

\section{Two Lax representations}
\setcounter{equation}{0}
\subsection{$2\times 2$ versus $n\times n$ representations}
So far we dealt with ``lattice" Lax
representations \cite{FT}, for the $SL(p)$ groups formulated in
terms of $p\times p$ matrices whose product along the chain --
transfer-matrix -- gives rise to the spectral curve. However, it is often
needed another type of the Lax representations given by $n\times n$ matrices
that is adequate for the (spin) Calogero (Toda) systems (and, more generally,
Hitchin systems \cite{Hitsys,nikita2,rub}). In this case, the Lax operator
itself generates the spectral curve, while no analog of the transfer-matrix
is known.

However, it sometimes happens that the same integrable system can possess
{\it both} Lax representations resulting in the same spectral
curve and different but equivalent phase space descriptions for the system.
In the brane language it can be formulated as the equivalence after the
rotation of two brane configurations. An example of this situation will be
considered in the next section. It appears that this issue can be formulated
in a relatively rigorous way.

In this and partly the next sections, we discuss interrelations between
different Lax representations and their interpretation in brane terms.

The simplest relevant example is the Toda chain which has the $2\times 2$
Lax representation (see s.{Toda})
\begin{equation}
L_{Toda}=\left(
\begin{array}{cc}
\lambda+p&e^q\\
-e^{-q}&0
\end{array}
\right)
\end{equation}
with the linear problem
(\ref{lproblem}) and the boundary conditions (\ref{pbc}).
This system can be reformulated in terms of $n\times n$ matrix. Indeed,
consider the two-component Baker-Akhiezer function $\Psi_n=
\left(\begin{array}{c}
\psi_n\\
\chi_n
\end{array}\right)$.
Then the linear problem (\ref{lproblem}) can be rewritten as
\begin{equation}
\psi_{n+1}-p_n\psi_n+e^{q_n-q_{n-1}}\psi_{n-1}=\lambda\psi_n,\ \ \ \
\chi_n=-e^{q_{n-1}}\psi_{n-1}
\end{equation}
and, along with the periodic
boundary conditions (\ref{pbc}) reduces to the linear problem
${\cal L}(w)\Phi=\lambda\Phi$ for the $n\times n$ Lax operator
\begin{equation}
\label{LaxTC}
{\cal L}(w) =
\left(\begin{array}{ccccc}
-p_1 & e^{{1\over 2}(q_2-q_1)} & 0 & &-
{1\over w}e^{{1\over 2}(q_{n}-q_1)}\\
e^{{1\over 2}(q_2-q_1)} & -p_2 & e^{{1\over 2}(q_3 - q_2)} & \ldots & 0\\
0 & e^{{1\over 2}(q_3-q_2)} & - p_3 & & 0 \\
 & & \ldots & & \\
-{w}e^{{1\over 2}(q_{n}-q_1)} & 0 & 0 & & -p_{n}
\end{array} \right)
\end{equation}
with the $n$-component Baker-Akhiezer function
$\Phi=\{e^{-q_n/2}\psi_n\}$. This leads
us to the spectral curve
\begin{equation}\label{hren}
\det_{n\times n}\left({\cal L}(w)-\lambda\right)=0
\end{equation}
which is still equivalent to the spectral curve (\ref{specurv}).

The symmetry of two different representations of Lax operator by
$2 \times 2$ or $N \times N$ matrices can be clearly reinterpreted
in string theory language.

From the brane pictures point of view the two Lax representations
are nothing but the choice of parametri\-za\-ti\-on of the fivebrane
worldvolume $\Sigma$. Of two holomorphic coordinates (which are
Cartan elements of scalars) $\la = x^4 + i x^5$ and $s = x^6 + i x^{10}$,
\footnote{Equivalently, $s= x^6 + i x^9$ in the Type IIB language.} we can
regard any we wish as a spectral parameter. In the ``standard" $2 \times 2$
representation, $w=\exp{-s \over R}$ is a spectral parameter, and the spin
chain is the appropriate integrable system. The corresponding typical brane
picture is depicted in Fig.4. Another possibility is to choose
$\la$ as a spectral parameter. It leads to the $n \times n$ representation.

Geometrical image of this phenomenon is the base-fiber
symmetry of the Calabi-Yau threefold \cite{V1}. As it was already mentioned
earlier, in the ``standard" $2 \times 2$ language $w$ parametrizes
the base $A_{m}$ singularity and counts the number $m$ of gauge factors.
$\la$ is a coordinate in the fiber. If each coefficient at $w^i$ is
a polynomial of degree $n$, we deal with the $SU(n)^m$ gauge theory with
$n$ extra fundamentals at each end of the chain of the gauge groups.
The duality
$SU(n)^m \leftrightarrow SU(m+1)^{n-1}$ proposed in \cite{V1} is
nothing but the $m+1 \times m+1 \leftrightarrow n \times n$ symmetry
of the Lax representations.
Under this symmetry, the Newton polygon (Fig.3) reflects so that
axis $i$ interchanges with axis $j$.

Note that the holomorphic differential
\be
dS = \la {dw \over w} = \la ds \approx - s d \la
\ee
also suggests this symmetry.

\subsection{Sixbranes versus semi-infinite branes}

Below we argue that the representation of the fundamental
matter by the D6 branes and the semi-infinite D4 brane has different
interpretation in terms of integrable systems. Namely, we would like
to propose that the D6-branes are natural in the $2\times 2$
representation, while the semi-infinite D4 branes in the $n\times n$ one.

We start with making use of the effect
(first found by Hanany and Witten \cite{HW}) of creating
a new brane when two other branes pass through each other.

\begin{figure}
\epsfxsize 400pt
\epsffile{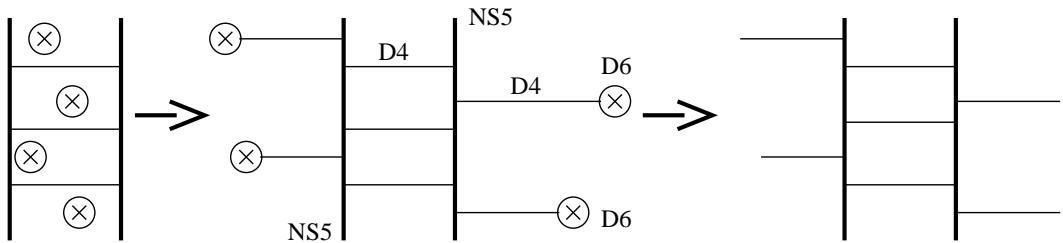}
\caption{Different representations of fundamental matter first
by inserting sixbranes, and then transformation of them into
semi-infinite fourbranes}
\end{figure}

In the paper, the authors considered the
three dimensional theory in the common longitudinal directions of
the branes:

\begin{equation} \ \ \ \left\{
\begin{array}{c|cccccccccc}
      & 0& 1& 2& 3& 4& 5& 6& 7& 8& 9\cr
NS5   & +& +& +& +& +& +& -& -& -& -\cr
D3    & +& +& +& -& -& -& +& -& -& -\cr
D5    & +& +& +& -& -& -& -& +& +& +
\end{array}\right.
\end{equation}

By T-duality in the third direction, this setup gets
mapped into Type IIA brane picture \cite{W} described above (\ref{tabl})
so that the statement of
creating a new D3-brane when D5-brane passes through NS5-brane
transforms into the following one: moving D6-brane through NS5-brane
creates a new D4-brane stretched between them.

As it was also mentioned in \cite{HW,W}, the $x^6$-coordinate of the sixbranes
apparently does not
play any role, at least, in the low-energy physics we are interested in.
Therefore, one can take these D6-branes to infinity, stretching a new
D4-brane each time we pass through a NS5-brane. Moreover, we can move
every particular sixbrane independently to the left or to the right
up to our choice. The final configuration, shown in Fig.2, includes
the original setup of fivebranes and fourbranes and very long
fourbranes (with a sixbrane attached to the end of each D4-brane) instead
of the sixbranes. Then, if the sixbranes (endpoints of the D4-branes) are
quite far from the picture we are dealing with, we can replace this setup of
branes by the same without any sixbranes, since they do not affect the
physics. For example, the beta-functions of the gauge theories remain the
same, since the presence of the D6-branes at infinity does not change
the relative bending of the fivebranes. In other words, it means that we are
going to the infinity of $Q_4$ which is Asymptotically Locally Flat.

The use of this phenomenon can be found in other columns of Fig.1.
For instance, the existence of two different brane pictures which
reveal the same
low-energy physics suggests the existence of two different integrable
counterparts. Really, the appropriate counterparts for the models with
sixbranes are spin chains, while the semi-infinite fourbranes find their
description in deformations of the Toda system by nontrivial boundary
conditions suggested by I.Krichever and D.Phong \cite{KP}. Though at the
low-energy level both systems are equivalent, the difference must appear in
higher terms in the field theory, or higher times in the
corresponding integrable systems. This also responds the question
about the role of the $x^6$-coordinates of the sixbranes.

Let us more comment the same point from a slightly different view.

In the $2\times 2$ representation, the spectral curve is
(\ref{scToda}). This spectral curve describes the pure gauge system obtained
by degeneration of the (D6-branes + pure gauge) theory (spin magnet).
Therefore, it is natural to identify the variable $w$ in (\ref{scToda}) with
the variables $y$ or $z$ in \cite{W}. These latter can be considered as
describing one of the complex structures of the multi-Taub-NUT solutions
\be
yz=\prod(\lambda-e_i)
\ee
where $e_i$ are $(x^4+ix^5)$-coordinates of the sixbranes.
It is of crucial importance that $y$ and $z$ never go to
infinity at finite $\lambda$ and, thus, describe the brane configuration
that have only points at finite distances whenever $\lambda$ is finite.

In contrast to the $2\times 2$ representation, in the $n\times n$
representation the spectral curve has the form (\ref{hren})
\be\label{hren1}
w+{1\over w}=P(\lambda)
\ee
and is obtained from the Lax operator (\ref{LaxTC}) given on a cylinder
(double-punctured sphere). Let us note that the curves (\ref{scToda}) and
(\ref{hren1}) are identical only if $w\ne 0$. If, however, one considers the
projectivization of these spectral curves, they will become identical, with
(\ref{scToda}) and (\ref{hren1}) corresponding to different maps. From
the point of view of the Toda theory, this merely corresponds to different
choices of gradation in Lax operator.

It is clear that the variable $w$  lives on the cylinder with radius $R$
which is identified
with the radius of the compactified eleventh dimension in M theory \cite{W}
\be
w=t=e^{-{x^6+ix^{10}\over R}}
\ee
The zero values of $w=t$ describe infinitely far points of the brane
configuration. If now one adds the matter hypermultiplets in the fundamental
representation, the spectral curve takes the form
\be\label{hren2}
w+{Q(\lambda)\over w}=P(\lambda)
\ee
equivalent to (\ref{scsl2m}) unless $w=0$. However,
there will be the points on the curve with zero values of $w=t$ (i.e.
infinite values of physical coordinates) located at finite $\lambda$'s. These
points describe the semi-infinite D4-branes \cite{W} and can be thought of as
infinitely long handles, or punctures on the Riemann surface that correspond
to the matter fundamental hypermultiplets in the standard picture
\cite{WDVV}.

The curve (\ref{hren2}) can be immediately obtained from the
$n\times n$ representation by the simple deformation of the Lax operator
(\ref{LaxTC}) just filling up the first and the last rows of the matrix
(\ref{LaxTC}) with any constant (independent on $\lambda$ and
proportional to $w^{-1}$ and to $w$ respectively) entries \cite{KP}. This Lax
representation can be just obtained as already mentioned non-local boundary
condition in the Toda chain \cite{KP}. Certainly, one can equally add
non-zero entries to the first and the last columns. These two
possibilities correspond to attaching the semi-infinite branes to the right or
to the left NS5-branes respectively. We return to this construction again in
the next subsection.

In a word, we would associate $2\times 2$ spin chains with the system with
matter fundamentals realized via inserting the sixbranes. On the contrary,
the $n\times n$ representation is associated with the realization of the
matter hypermultiplets via the semi-infinite branes.

There is another argument in favor of this identification that is related to
the integrable system $\leftrightarrow$
brane configuration correspondence. Indeed, it is natural to consider the
influence of the semi-infinite branes as a boundary effect like the
emergence of $Q(\lambda)$ in the paper \cite{KP} is due to the non-local
boundary conditions. At the same time, the sixbranes are more naturally
associated with the ``local" perturbations of the D4-branes suspended between
NS5-branes, i.e.  with additional degrees of freedom in the spin chain. This
nicely fits the recent observation \cite{Ooguri} that the semi-infinite
D4-branes can be more naturally attributed to the meson not quark degrees of
freedom.

\subsection{$n\times n$ representation in the higher rank magnets}
In this subsection we discuss a way to obtain $n\times n$ representations by
degenerating the $SL(p)$ magnet. On this way, we reproduce the Lax operator
of \cite{KP} that corresponds to bringing to infinity all but the very left
two NS-branes. Then, the resulting system looks as the $SU(n)$ SYM theory
with massive hypermultiplets realized via semi-infinite branes attached to the
right NS-brane.

To begin with, let us note that, starting from eq.(\ref{hren}), one can
maximally reproduce the curve (\ref{hren1}) with polynomials $P(\lambda)$ of
degree $n$ and $Q(\lambda)$ of degree $2n-2$. Thus, within the $n\times n$
representation, one can not reproduce the reference system (that requires the
degree $2n$ of the polynomial $Q(\lambda)$) but can approach to it as close
as possible\footnote{Note that, following \cite{KP}, one should consider
$2n\times 2n$ Lax matrix to include the UV-finite case (i.e. to
obtain $Q(\lambda)$ of degree $2n$) into the play. Thus doing, one is led
to consider also any $pn\times pn$ Lax matrices -- the result we can not
interpret.}.

Now let us consider the pure gauge limit of the $SL(p)$ magnet with the Lax
operator with only one diagonal entry containing $\lambda$. From the results
of s.4.3 we can conclude that it has the form
\begin{equation}
\label{LaxL}
{L}_{p\times p}(\lambda) =
\left(\begin{array}{cccccc}
\lambda+S_{0}^{(1)}
& S_{-}^{(1)} & S_{-}^{(12)} & S_{-}^{(13)}&\ldots
&S_{-}^{(1,p-1)}\\
S_{+}^{(1)} & &&&&\\
S_{+}^{(12)} & &&&&\\
S_{+}^{(13)} & &&\hat A&&\\
\vdots & &&&&\\
S_{+}^{(1,p-1)}&&&&&
\end{array} \right)
\end{equation}
where $\hat A$ is an ordinary constant $p\times p$ matrix of rank $p-2$
(thus, with zero determinant), $\det L_{p\times p}=1$ and the Poisson
brackets are
\be
\left\{S_{\pm}^{(1i)},S_0^{(1)}\right\}=\pm S_{\pm}^{(1i)},\ \ \ \
\left\{S_{\pm}^{(1i)},S_{\mp}^{(1j)}\right\}=A_{j+1,i+1},\ \ i,j=0,1,...,
p-1,\ \ S_{\pm}^{(1,0)}\equiv S_{\pm}^{(1)}
\ee
This Lax operator leads to the spectral curve
\be\label{hh1}
w^p+P^{(1)}_n(\lambda)w^{p-1}+\ldots+P^{(k)}_n(\lambda)w^{p-k}+\ldots+1=0
\ee
and, therefore, has no $n\times n$ representation (in accordance with the
argument in the beginning of this subsection). However, one can restrict the
matrix $\hat A$ further so that the polynomial $P^{(k)}(\lambda)$ in
(\ref{hh1}) would be of degree $n-2k$. This system already has an $n\times n$
representation. It can be reached, for instance, in the case of $p=3$ by the
choice
\be
\hat A=\pmatrix{1&-1\cr 1&-1}
\ee
in the case of $p=4$ by the choice
\be
\hat A=\pmatrix{1&-1&1\cr 1&-1&1\cr 1&-1&0}
\ee
etc. In the Appendix we consider how it happens in detail in the $p=3$ case.

The Lax operator in this $n\times n$ representation has the following
structure (for $n>p$):
\begin{equation}
\label{hh2}
{\cal L}_{n\times n}(w) =
\left(\begin{array}{ccccccc}
S_{0,1}^{(1)} & 1 & \ldots && w^{-1}&\ldots &\ast\cdot w^{-1}\\
\ast &S_{0,2}^{(1)}&1&\ldots&w^{-1}&\ldots &\ast\cdot w^{-1}\\
\ast&\ast&S_{0,3}^{(1)}&1&\ldots&\ldots &\ast\cdot w^{-1}\\
&& &\ddots&&&\\
&& \ddots&&&&\\
&& &&&&\\
0&\ldots &\ast&\ldots&*&S_{0,p-1}^{(1)}&1\\
w&0 &\ldots&\ast&\ldots&*&S_{0,p}^{(1)}\\
\end{array} \right)
\end{equation}
so that the $p-1$ right corner diagonals are filled up by non-zero entries
proportional to ${1\over w}$ and $p-1$ diagonals below the main one are also
non-zero. In formula (\ref{hh2}) we use the asterisque to denote the non-zero
entries which we do not write down manifestly for the sake of simplicity.

Here we use the normalization that slightly differs from (\ref{LaxTC}). This
Lax operator, after inserting into (\ref{hren}), indeed, leads to the
spectral curve ($n>p$):
\be
w^p+P^{(1)}_n(\lambda)w^{p-1}+P^{(2)}_{n-2}(\lambda)w^{p-2}+\ldots+1=0
\ee
Moreover, for $n=p$, one even gets the curve
\be
w^p+P^{(1)}_n(\lambda)w^{p-1}+P^{(2)}_{n-1}(\lambda)w^{p-2}+\ldots+1=0
\ee
that is the maximal possible curve following from (\ref{hren}), at given $p$
and $n$.

If now one brings all but the very left two NS-branes to infinity, the
system looks as an $SU(n)$ system with matter hypermultiplets realized via
the semi-infinite branes, the Lax operator (\ref{hh2}) results into that of
\cite{KP}, and the spectral curve takes the form (\ref{hren1}).  Thus, the
construction of \cite{KP}, indeed, corresponds to the realization of the
matter hypermultiplets via the semi-infinite branes.

\subsection{Two Lax representations: general comments}

Note that the two ``dual" Lax representations has been presented in literature
\cite{harnad} in more invariant terms in the Gaudin limit of the Toda chain
system. In the $2\times 2$ representation, this limit corresponds to the
replacement $L_{2\times 2}\to 1+\epsilon L_G$ with small $\epsilon$ and leads
to the {\it linear} Poisson brackets, instead of (\ref{quadr-r}). Following
\cite{FT}, one can associate this replacement with transition to
the continuum limit.
In terms of the $n\times n$ Lax matrices, this limit corresponds to the
rational limit of the trigonometric Toda $r$-matrix.

The relevant description of the Gaudin Lax representations can be
presented via the embedding of this finite dimensional integrable system into
rational coadjoint orbits of loop algebras. It is based on a family of the
momentum maps from the space
$M= \{F,G \in M^{n \times r}\}$ of pairs of
$n \times r$ rectangular matrices, with a natural symplectic structure
\be
\omega = \tr(dF \wedge dG^T)
\ee
to the dual of the positive half of the loop algebra $\wt{gl(r)}$.
This algebra is defined as the semi-infinite formal loop algebra on $gl(r)$
that consists of elements $X(\lambda)=\sum^m_{i=-\infty}X_i\lambda^i$ with
$X_i\in gl(r)$. Algebra $\wt{gl(r)}$ as the vector space has a natural
decomposition
\be
\wt{gl(r)}=\wt{gl(r)}^+\oplus \wt{gl(r)}^-
\ee
into the spaces of $r\times r$ matrix polynomials and strictly negative
formal power seria in $\wt{gl(r)}$ respectively. We define the pairing
between these two factors
\be\label{h1}
\left<X(\lambda),Y(\lambda)\right>=\Tr\left(X(\lambda)Y(\lambda)
\right)_{-1}
\ee
where the subscript ``-1" refers to the coefficient of $\lambda^{-1}$.

Now we fix an $n\times n$ matrix $A$ whose spectrum is completely inside a
disk $D$ and define the group $\wt{GL(r)}^+$ of $GL(r)$-valued functions of
$\lambda$ that are holomorphic inside $D$.

Each this matrix $A$ defines a symplectic action of $\wt{GL(r)}^+$ on $M$:
\be
g(\lambda):\ \left(F,G\right)\to\left(F_g,G_g\right),\ \ \ \
g(\lambda)\in \wt{GL(r)}^+\\
\left(A-\lambda\right)^{-1}Fg^{-1}(\lambda)=
\left(A-\lambda\right)^{-1}F_g+F_{hol},\\
g(\lambda)G^T\left(A-\lambda\right)^{-1}=G^T_g\left(A-\lambda\right)^{-1}
+G^T_{hol}
\ee
where $F_{hol}$, $G^T_{hol}$ are holomorphic in $D$.

One can easily check using (\ref{h1}) \cite{harnad} that this symplectic
action is Hamiltonian, with moment map $J_A:\ M\to \left(\wt{gl(r)}^+
\right)^*$ defined by
\be
J_A(F,G)=-G^T\left(A-\lambda\right)^{-1}F
\ee

Absolutely analogously to this scheme, one can consider the loop algebra
$\wt{gl(r)}^+$ and its group $\wt{GL(r)}^+$ and introduce a fixed $r\times r$
matrix $Y\in gl(r)$ that define the action
\be
h(w):\ \left(F,G\right)\to\left(F_h,G_h\right),\ \ \ \
h(w)\in \wt{GL(n)}^+\\
h(w) F\left(Y-w\right)^{-1}=
F_h\left(Y-w\right)^{-1}+\bar F_{hol},\\
\left(Y-w\right)^{-1}G^Th^{-1}=\left(Y-w\right)^{-1}G^T_h
+\bar G^T_{hol}
\ee
where $\bar F_{hol}$, $\bar G^T_{hol}$ are holomorphic inside disk $\bar D$
containing the spectrum $Y$. As above, this action is Hamiltonian, with
moment map $J_Y:\ M\to \left(\wt{gl(n)}^+\right)^*$ defined by
\be
J_Y(F,G)=F\left(Y-w\right)^{-1}G^T
\ee

Now let us define two matrices
\be
L(\lambda)\equiv Y+G^T\left(A-\lambda\right)^{-1}F,\\
{\cal L}(w)\equiv A+F\left(Y-w\right)^{-1}G^T
\ee
Using the Adler-Kostant-Symes theorem, one now proves that these two matrices
are Lax matrices \cite{harnad}. Each of them is the Lax matrix of the
classical Gaudin system. They can be also considered as solutions to the
moment map equation of the Hitchin system on the sphere with $n$ or,
respectively, $r$ marked points so that $Y$ and $A$ define the monodromy of
infinity.

The equivalence of these two Lax representations is defined by the fact that
they define bi-rationally equivalent spectral curves. This follows from the
identity
\be
\det\left(A-\lambda\right)\det\left(Y+G^T\left(A-\lambda\right)^{-1} F-w
\right)=
\det\left(Y-w\right)\det\left(A+F\left(Y-w\right)^{-1}G^T-\lambda
\right)
\ee

Note that the construction for the Toda case looks more involved, since,
in the $2\times 2$ formalism, it is described by the quadratic Poisson
brackets. This results in the product of the Lax operators along the chain
(polynomial in $\lambda$ of degree $n$).
Actually we have to perform the transition from the algebra to the group
level.
Put it differently, this is
the product over marked points, instead of sum that could be reproduced by
the larger matrix still linear in the spectral parameter. Analogously,
$n\times n$ Toda Lax operator has {\it linear} Poisson brackets with {\it
trigonometric} $r$-matrix. This can be again reproduced by the product of two
matrices (leading to a polynomial quadratic in $w$) instead of sum over the
two marked points. Unfortunately, there is no technique developed for the
products of matrices. It looks slightly similar to the averaging procedure
proposed in \cite{FR}, which is immediate for the linear Poisson brackets and
is quite involved for the quadratic ones.

Let us also note that the natural analog of matrices $F$ and $G$ in the Toda
case would be the $n\times 2$ matrix with the two columns made of
$\{\psi_n\}$ and $\{\chi_n\}$.

Let us still say some words of the algebraic structures in the Gaudin system
obtained from the Toda chain which corresponds to the above construction at
$r=2$. In the $n\times n$ representation, the Lax operator is defined as the
solution of the moment map equation for the holomorphic $SU(n)$ connection on
the sphere with two marked points. The Lax operator in this
representation contains momentum variables as the diagonal entries. This
can be interpreted in the framework of the consideration above as the
monodromy at infinity. Turn now to the $2\times 2$ representation.
Its structure clearly indicates that now momenta play
the role of the positions of the marked points on the punctured sphere with
the holomorphic coordinate $\lambda$ that was the eigenvalue of the Lax
operator in the first representation. The picture can be clearly generalized
to the Gaudin system with arbitrary number of marked points in the first
representation.

Returning to the
brane interpretation behind the dual Lax representations, it can be viewed
as follows.
In the Type IIA, in one representation we ``look from" the diagonal of the
brane picture which
gives rise to the $2\times 2$ representation.
It is this look that admits the
interpretation in terms of an ``effective $n$-monopole".
In the other, $n \times n$ representation we take another look
and obtain $SU(n)$ gauge group from the parallel D4-branes. The
D0-branes provide the structure with a few monopoles
in the $SU(n)$ theory leading to the $n\times n$ representation.
In the IIB picture, we have $U(1)$ gauge field on NS5 brane and a
possible interpretation of $2\times 2$ representation is
discussed in \cite{HW}.
The other, $n\times n$ interpretation, in this case, comes from the wrapped
branes around the compact surface.

\section{Monopole $\leftrightarrow$ spin chain correspondence}
\setcounter{equation}{0}
It was already noted by different authors \cite{sat,Dia,HW,Gor} that
monopole moduli spaces play an important role in
the 4d SUSY gauge theories. The evident reason for the appearance
of monopoles in this setup is the hyper-K\"ahler structure of the
monopole moduli spaces which also can be manifested for the
moduli space of vacua in SYM theories as well as for the phase space of
the integrable systems under consideration. The useful
approach to the description of the monopole moduli space is to
introduce monopole spectral curve and look at the moduli space of such
curves \cite{Hit2}. It was shown \cite{Nahm,Hit} that the monopole
spectral curve is related to the spectral curve of the so-called Nahm
equations giving an infinite-dimensional counterpart of the ADHM
construction \cite{Nahm,Hit}. It was P.Sutcliffe who first used the Nahm
equations in the context of the Seiberg-Witten theories and noted that the
solutions of the periodic Toda chain are naturally identified with the $C_p$
cyclic $SU(2)$ monopoles of charge $p$.

It is worth noting that Nahm approach suggests an interpretation of the
equations of motion of our integrable systems. Indeed, it was mentioned in
\cite{Dia} that the Nahm equations (and, therefore, the equations of motion
of integrable system) are interpreted as the condition of the BPS saturation
which keeps some supersymmetries. This means that the demand for the
vacuum state in the proper $\sigma$ model to be supersymmetric gives
rise to the equations of motion. Note also that the fermions in the auxiliary
Lax linear problem (Baker-Akhiezer function) would be related to the fermion
zero modes in a monopole background.

In this section we demonstrate that such a correspondence between monopoles
and integrable chains can be pushed further and state that the solutions of
the spin chains can {\it exhaust all} the solutions to the Nahm equations. As
a byproduct, we obtain a natural generalization of these equations.
Physically the proposed correspondence implies a duality of some brane
pictures. This can be understood in the framework of the Diaconescu
construction \cite{Dia}, who generalized the analogous picture for instantons
\cite{douglas}.

\subsection{$SU(2)$-monopole versus spin chains}

We start with the $SU(2)$ monopole\footnote{Hereafter, speaking of monopole
we mean BPS monopole.} with charge $p$. It can be described by the Nahm
equations \cite{Hit}
that are three equations for three $p\times p$ matrices
${\cal T}_i(t)$, $i=1,2,3$
\be\label{Nahm1}
{d{\cal T}_i\over dt}={1\over 2}\epsilon_{ijk}
\left[{\cal T}_j,{\cal T}_k\right]
\ee
given on the real segment $t\in [-1,1]$ and
regular on the interval $t\in (-1,1)$. These matrices
satisfy additional restrictions
\be\label{C12}
{\cal T}_i(t)=-{\cal T}^{\dag}_i(t),\ \ \ {\cal T}_i(t)=
{\cal T}_i^{\hbox{tr}}(-t)
\ee
and boundary conditions
\be\label{C34}
{\cal T}_i(t)\stackrel{t\to 1}{\sim} {T_i^{(1)}\over t-1},\ \ \ \
{\cal T}_i(t)\stackrel{t\to -1}{\sim} {T_i^{(-1)}\over t+1}
\ee
where ${\cal T}_i^{(1,-1)}$ form an irreducible $p$-dimensional representation
of $SU(2)$. The numbers 1, -1 are just the chosen normalization
of the asymptotics of the two Higgs field components at infinity. The physical
monopole fields are constructed from the zero modes of the linear first order
differential operator in the background field of the Nahm matrices
${\cal T}_i(t)$ \cite{Nahm,CG}.

The three Nahm equations (\ref{Nahm1}) can be rewritten as the single
equation depending on the spectral parameter. For doing this, one needs to
introduce the Lax operator
\be\label{NahmLax}
L_N(\lambda)={\cal T}_++{\cal T}_0\lambda+{\cal T}_-\lambda^2
\ee
where we denoted ${\cal T}_{\pm}={\cal T}_1\pm i{\cal T}_2$,
${\cal T}_0=-2i{\cal T}_3$.  Then, (\ref{Nahm1}) can be written as the single
Lax equation
\be\label{Nahm2}
{dL_N(\lambda)\over dt}=\left[A(\lambda),L_N(\lambda)\right],\ \ \ \
A(\lambda)={1\over 2}\left({\cal T}_+\lambda^{-1}-{\cal T}_-\lambda\right)
\ee
With the Lax representation, one can consider the isospectral
problem and construct conservation laws from the spectral curve:
\be
\det_{n\times n}(L_N(\lambda)+w)=0
\ee
This curve has the form \cite{Hit2,Hit}
\be\label{scNahm}
w^p+q_1(\lambda)w^{p-1}+...+q_{k}(\lambda)w^{p-k}+...+q_p(\lambda)=0
\ee
where $q_k(\lambda)$ is a polynomial of degree $2k$, and,
therefore, the genus of the curve is $(p-1)^2$ (in accordance with
(\ref{genus}). This curve is an
algebraic curve in the space of the oriented geodesics in ${\bf R}^3$
isomorphic to the complex tangent bundle $T{\bf P}^1$ of the Riemann sphere
${\bf P}^1({\bf C})$ \cite{Hit2}. If $x$ is the local coordinate on this
sphere, $(w,\lambda)$ are the standard local coordinates on $T{\bf P}^1$
given by $(w,\lambda)\to w{\partial\over\partial x}|_{x=\lambda}$. Now
conditions (\ref{C12}) and (\ref{C34}) can be rewritten in terms of the
spectral curve $S$ \cite{Hit2} as the reality condition
\be\label{B2}
q_k(\lambda)=(-1)^k\lambda^{2k}\overline{q_k(-\lambda^{-1})}
\ee
and the following restrictions:
\be\label{B3}
L^2 \hbox{ is trivial on the curve, } L(p-1) \hbox{ is real}
\ee
\be\label{B4}
H^0(S,L^{\omega}(p-2))=0 \hbox{ for } \omega\in(-1,1)
\ee
where $L^{\omega}$ is the linear holomorphic bundle over $T{\bf P}^1$ given
by the transition function $\exp(-\omega w/\lambda)$ defined on the domain
$\lambda\ne 0,\infty$, and $L(p)=L\otimes {\cal O}(p)$.

In what follows we establish the ``formal" correspondence between Nahm
equations and spin chains ignoring the additional requirements (\ref{C12}),
(\ref{C34}) or (\ref{B2})-(\ref{B4}) that just restrict the admissible space
of solutions. Thus, we concentrate on the general integrable properties of
the Nahm system with no respect to the boundary conditions. In fact, these
additional requirements are quite restrictive, since the monopole moduli
space has the dimension $4p$ increasing with $p$ linearly in contrast to the
genus $(p-1)^2$ and the quadratic grow of the dimension of the moduli space
of the general curves (\ref{scNahm}).

First of all, let us comment
interpretation of Toda integrable equations as a particular Nahm equation
(for the symmetric monopole configuration) \cite{sat,Gor}. The key property
here is that both coordinates and momenta of the integrable systems can be
expressed in the monopole terms. More concretely, the Lax matrix of the
integrable system $L(\lambda)$ can be written as a combination of the Nahm
matrices for $n$ symmetric $SU(2)$ monopoles (see below). However, first we
see how the monopoles are getting involved in the picture. Actually
as we discussed in section 2 we would
like to relate the coordinates of Toda particles with the positions of the
well separated monopoles in $x^{6}$-coordinate. The monopoles are
localized  on D4-branes, one per each brane, so we can attribute the same
coordinate $p_{i}=x^{4+i5,D4i}$ to the monopole localized on the i-th D4
brane. Note that this is in agreement with the ``inversion formula" of the
Nahm construction \cite{CG}
\be\label{inverse}
{\cal T}_{i}^{km}(t)=\int d^{3}x \Psi^{+,k}(x,t) x_{i}\Psi^{m}(x,t)
\ee
which gives the Nahm matrices ${\cal T}_i(t)$ as the matrices of the
coordinates in the Heisenberg picture in the space of the fermionic zero
modes $\Psi^{k},(k=1,..,n)$ in the monopole background. The relevant
Nahm matrices provide the momenta $p_{i}$ in the Cartan sector, and
coordinates $e^{iq_{n}}$ for positive and negative roots
\be\label{NahmToda}
{\cal T}_{1}=\frac{i}{2}\sum_{j=1}e^{iq_{j}}(E_{+j}+E_{-j}) ; \ \ \ \
{\cal T}_{2}=-\sum_{j=1}e^{iq_{j}}(E_{+j}-E_{-j})  ;           \ \ \ \
{\cal T}_{3}=\frac{i}{2}\sum_{j}p_{j}H_{j}
\ee
where $E_{\pm j}$ are the generators corresponding to the simple and
longest roots of the affine algebra $\widehat{SL(n)}$ and $H_j$ are its
Cartan generators. Note that the formulae above can be also interpreted
in terms of the fundamental monopole of $SU(n)$.

Now let us return to the general monopole (Nahm) system and,
first, let us note that the spectral curve (\ref{scNahm}) coincides with the
spectral curve of the $SL(p)$ spin chain at two sites (\ref{sc}). In
particular, one can easily get from the genus formula (\ref{gf}) for $n=2$
the genus $(p-1)^2$ that is well-known answer for the monopole
spectral curve \cite{Hit2}. Therefore, it is naturally to suggest that these
two systems are equivalent. In fact, the identity of the spectral curves is
still not sufficient to identify two systems. Thus, one needs to check the
identity of the Lax representations. As soon as there is a gauge
transformation that connects two Lax operators, the two systems are
equivalent possessing the same integrals of motion. In this concrete
situation should connect the Nahm Lax operator (\ref{NahmLax}) and the Lax
operator (\ref{Lax}) of the $SL(p)$ spin chain.  In fact, we assert that the
2-site transfer matrix
\be\label{2sT}
T_2(\lambda)=L_2(\lambda)L_1(\lambda)
\ee
reproduces (\ref{NahmLax}). Indeed, the Nahm Lax operator is defined just as
an arbitrary matrix polynomial quadratic in $\lambda$, i.e. all the three
coefficients ($p\times p$ matrices) are arbitrary. On the other hand, the
2-site transfer matrix (\ref{2sT}) is the product of two arbitrary {\it
linear} matrix polynomials with {\it the unit} leading coefficient. This can
give us only two arbitrary $p\times p$ matrices. The third one comes from the
possibility to multiply spin chain Lax operator by any constant matrix
without changing the conservation laws (remind that this is the case, since
the $r$-matrix (\ref{rg}) is the permutation operator and, therefore,
commutes with the expressions like $U\times U$).

\begin{figure}
\epsfxsize 400pt
\epsffile{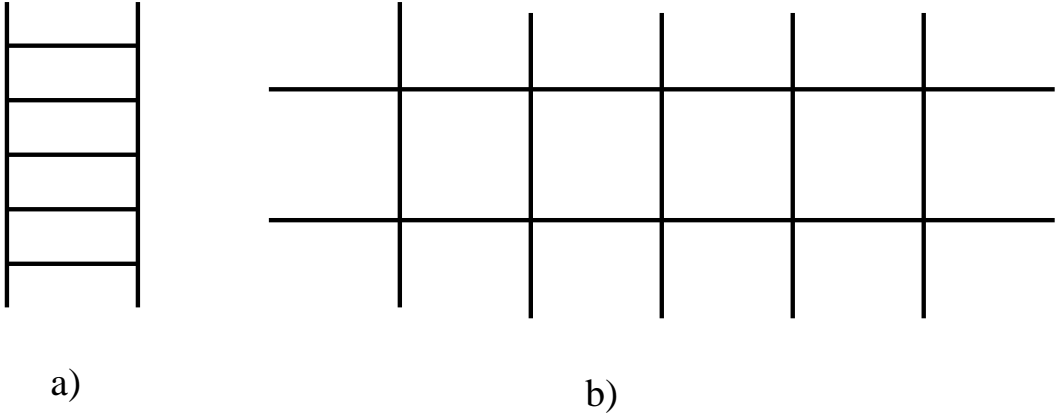}
\caption{Two looks at the Toda chain: explanation of the embedding
Toda $\hookrightarrow$ Nahm}
\end{figure}

With the established correspondence between $SL(p)$ spin magnet on two sites
and the Nahm equations, one can immediately explain the embedding
(\ref{NahmToda}) of the Toda Lax operator into the Nahm one. Indeed, let us
look at the $SU(p)$ pure gauge supersymmetric theory that is described by the
brane configuration of Fig.5a corresponding to the $SL(2$ Toda chain ($SL(2)$
magnet) given on $p$ sites. This configuration can be equally viewed as that
depicted in Fig.5b. This latter point of view implies the interpretation of
the same system as a specific case of the $SL(p)$ spin magnet given on two
sites. This is as stated above equivalent to the Nahm system. Since the
$SL(p)$ magnet (= Nahm system) corresponding to the brane configuration of
Fig.5b is not of the general form, one should expect that the Toda Lax
operator (Fig.5a) is only {\it embedded} into the Nahm Lax operator. This is,
indeed, realized in formula (\ref{NahmToda}).

Thus, we have proved that the $SU(2)$ p-charged Nahm system is really
equivalent
to the 2-site $SL(p)$ spin chain. This hints that the spin chain at
many sites is equivalent to the higher group Nahm systems. In the next
subsections we check this idea.

\subsection{$SU(n)$-monopole, spin chains and generalized Nahm/Diaconescu
construction}

We start with the description of the $SU(p)$ Nahm construction following
\cite{Hurt,Dia}. In this case, we fix the large-distance asymptotics of the
$n$ components of the Higgs field to be $\mu_i$, $i=1,...,n$,
$\sum_i\mu_i=0$. It naturally divides the domain where the Nahm system is
given onto $n-1$ intervals $(\mu_1,mu_2)$, ..., $(\mu_{n-1},\mu_n)$. On each
$a$-th interval there is defined the Nahm system for the three
$p_k\times p_k$ matrices ${\cal T}_{i,a}(t)$
\be\label{Nahmp1}
{d{\cal T}_{i,a}\over dt}=
{1\over 2}\epsilon_{ijk}\left[{\cal T}_{j,a},{\cal T}_{k,a}\right]
\ee
Thus, $SU(n)$ monopole is defined by the set of charges (topological numbers)
$p_1,...,p_{n-1}$.

The system (\ref{Nahmp1}) requires some boundary and matching
conditions. The details can be found in \cite{Hurt}, here we only describe
the matching condition at the boundary of two intervals $\mu_a$ for
the Nahm matrices with some $p_a$ and $p_{a-1}$. Let us suppose for
definiteness that $p_a\ge p_{a-1}$. In the case of equality, all matrices
${\cal T}_{i,a}(t)$ and ${\cal T}_{i,a-1}(t)$ are analytic nearby $\mu_a$
although the condition ${\cal T}_{i,a}(t)={\cal T}_{i,a-1}(t)$ around $\mu_a$
may be not fulfilled.  In the case of strict inequality, ${\cal
T}_{i,a-1}(t)$ is analytic as $t\to\mu_a$, while ${\cal T}_{i,a}(t)$ has the
block form
\be
\left(
\begin{array}{cc}
{\cal T}_{i,a-1}(\mu_a)+O(t-\mu_a)& O((t-\mu_a)^{(p_a-p_{a-1}-1)/2})\\
O((t-\mu_a)^{(p_a-p_{a-1}-1)/2})& {T_{i,a}^{(\mu_a)}\over t-\mu_a}+O(1)
\end{array}
\right)
\ee
and $T_{i,a}^{(\mu_a)}$ realizes an irreducible representation of $SU(2)$.
These matching conditions have to be added by the boundary conditions
$p_0=p_n=0$.

Now let us turn to the spectral curve. From the described construction it is
clear that, by modulo boundary and matching conditions, the spectral curve is
the product of $n-1$ components, each of these being the spectral curve for
the $SU(2)$ monopole of charge $p_a$. The complete curve, therefore, is a
curve of the form (\ref{scNahm}) which, in its turn, is a
special degenerated curve of the form (\ref{sc}) with $p=\sum_a p_a$.
In the
spin chain language, this means that we are dealing with a very special
configuration of the $SL(p)$ magnet given on the $n$ sites.This
configuration is determined by ``the clasterization" of these $n$ sites into
2-site groups so that there is effectively given the $SL(p_a)$ magnet on each
of these groups, with specific matching conditions.

To understand the speciality of this configuration better, we need to
turn to the Diaconescu construction \cite{Dia} that could allow us to propose
a natural generalization of the Nahm construction to the generic $SL(p)$ spin
chain.

Looking at this construction, one can easily understand what is the specific
of the Nahm system that leads to a very peculiar magnet. This specific is
obliged to ``the locality" of the brane picture in the sense that the 1-branes
has their endpoints only on the neighbor 3-branes. This implies that the
corresponding spectral curve is just the product of the curves constructed
from each pair of the neighbor 3-branes. In terms of spectral curve,
this means that there preserved the asymptotic behavior at
large $w$ and $\lambda$, that is $w\sim\lambda^2$. Thus, appealing to the spin
chain interpretation proposed in this section, one can identify the 3-branes
with the sites of the spin chain, while the 1-branes gives the $SL(p)$ group,
$p$ being the full number of 1-branes. This leads us to the natural
conjecture that the generic $SL(p)$ spin chain corresponds to the generalized
Diaconescu picture where the 1-branes connects the very left and the very
right 3-branes passing through the others (inside) 3-branes. The
corresponding Nahm Lax operator is to coincide with the transfer matrix of
this spin chain (with Lax operators (\ref{Lax}) twisted by the constant
matrices) and can be represented in the form
$L_N(\lambda)=\sum_{i=0}^{n}{\cal T}_i\lambda^i$ with $n+1$ arbitrary
matrices ${\cal T}_i$.

\section{Multiple $\Lambda_{i}$-scales}
\setcounter{equation}{0}

In this section we introduce a new set of gauge models inspired by
some generalization of the periodical
Toda chain which contains many $\Lambda$ type parameters.
Let us emphasize from the very beginning that the possibility
to be discussed below is purely nonperturbative.
The main idea is to consider the correlated inhomogeneous
magnetic fluxes and the inhomogeneous
``D0 brane lattice" that can be obtained by a nontrivial double
scaling limit from the system described by the elliptic spin Calogero model.

Thus, we start with the spin generalization of the elliptic
Calogero system that can be the most directly viewed in the Hamiltonian
reduction picture \cite{nikita2,rub}. For the simplicity, in this section we
discuss the $GL(n)$ system, the reduction to the $SL(n)$ case being
trivial and immediate. The system is given by the Lax
operator \cite{Kri}
\be\label{Calspin}
L_{ij}(\lambda)=p_{i}\delta_{ij}+J_{ij}
exp\left(\left(x_{i}-x_{j}
\right)\frac{\lambda -\bar{\lambda}}{\tau-\bar{\tau}}\right)
\frac{\sigma(\lambda +x_{i}-x_{j})}{\sigma(x_{i}-x_{j})\sigma(\lambda)}
\ee
with the additional requirement $J_{ii}=0$.\footnote{Let
us note that generally these diagonal elements are just conserved
quantities. The choice made is consistent with the spin Calogero system
obtained from the dynamics of poles of the matrix KP system \cite{Kri}.}
Thus, we put $J_{ij}=(1-\delta_{ij})f_{ij}$, $f_{ii}=const$. This latter
condition is necessary for integrability of the system (see \cite{Babelon}).

The additional spin variables in the Lax operator (\ref{Calspin})
$f_{ij}$ satisfy the Poisson brackets
\be
\left\{f_{ij},f_{kl}\right\}=\delta_{jk}f_{il}-\delta_{il}f_{kj}
\ee
This symplectic form is generally degenerate and its restriction to the
symplectic leaves is related to fixing the rank of the matrix $f_{ij}$ to be
$l\le n$ so that the matrix can be presented in the form
\be
f_{ij}=\sum_{a=1}^l u_i^av_j^a
\ee
and corresponds to the generic complex orbit $\displaystyle{{GL(n,\C)\over
GL(n-l,\C) \times \left(\C^{\ast}\right)^l}}$. The standard spinless Calogero
system corresponds to the orbit describing the fundamental representation
$\displaystyle{{GL(n,\C)\over GL(n-1,\C) \times \left(\C^{\ast}\right)}}$,
i.e. $l=1$.

The corresponding spectral curve of the spin Calogero system, which generates
the conserved quantities, is given, as usual \cite{bv}, by the determinant
form and its genus calculated, say, via Riemann-Roch theorem is
$g=np-{p(p+1)\over 2}+1$.

We are not going to discuss here the elliptic system more but instead
consider the double scaling procedure that provides us
with a set of scales in the low-energy theory
after its degeneration to the rational limit.
Let us remind how it works in the spinless case. The steps are
as follows \cite{In}. First, we degenerate the bare spectral torus
$\tau\rightarrow i \infty$. Then, to provide the nearest neighbor
interaction, we introduce the homogeneous
coordinate ``lattice" with the large distance $\Delta$
between sites
\be
x_{j}=j\Delta+\phi_{j}
\ee
To see explicitely what kind of interaction emerges in this limit,
it suffices to look at the Weirstrass function giving
the Calogero potential, although the procedure can be easily repeated for the
Lax operator. The Calogero potential has form
\be\label{pot}
V(x_{ij})=g^2\sum_{i,j}\wp(x_{ij})
\ee
where $x_{ij}=x_{i}-x_{j}$, $g^2=f_{ij}f_{ji}$ is the
(coupling) constant that can be made non-depending on indices by
the proper gauge transformation of $f_{ij}$
and the Weierstrass function is defined as
\be\label{ws}
\wp(x_{ij})\equiv\partial^{2}\log \sigma(x_{ij}) =
\sum_{m} \frac{1}{\sinh^{2}(x_{ij}+m\tau)}
\ee
In the limit under consideration we introduce the renormalized coupling
constant $g=g_{0}exp(\Delta)$ so that $g_{0}$ will be ultimately nothing but
$\Lambda_{QCD}$. Now, choosing $\Delta\sim\tau$ and taking the limit $\tau\to
i\infty$, we see that only $m=0$ term survives in the sum (\ref{ws})
in the potential (\ref{pot}) so that the resulting potential reads as
\be
V_0(\phi_{i})=g_0^2\sum_{j=1}^{n-1}e^{\phi_{i+1}-\phi_i}
\ee
and describes open (non-periodic) Toda chain.

In order to get the periodic Toda chain, one needs to fix $\Delta={\tau\over
n}$. In this case, the $m=-1$ term in (\ref{ws})
also contributes into the sum and one
finally obtains the potential
\be
V_{TC}(\phi_{i})=V_0+g_0^2e^{\phi_{1}-\phi_n}
\ee
describing the periodic Toda chain.

Let us now consider the generic spin Calogero system with $l=n$.
In this case, there is a rich spectrum of possibilities so
that one would expect $n$ scale parameters instead
of a single one. Indeed, it is possible now to
introduce the inhomogeneous ``lattice"
with $n$ independent parameters which give rise to $n$ infrared
scales after the independent dimensional transmutation procedure in each
$U(1)$ factor. Quantitatively it corresponds to
the definition
\be
x_{i+1,i}=\Delta_{i}+\phi_{i+1,i}
\ee
with arbitrary $\Delta_{i}$ which are assumed to be large. The potential in
this case has the form
\be
V(x_{ij})=\sum_{i,j}f_{ij}f_{ji}\wp(x_{ij})
\ee

Consider now different possible limits. In the simplest
case we choose
\be\label{7.1}
f_{i+1,i}=f_{i,i+1}=g_{0,i}\exp(\Delta_{i})
\ee
with finite
$g_{0,i}$ assuming that no other $f_{i,j}$ have the same exponential factor.
This amounts to the open Toda chain with different coupling constants
$\Lambda_{i}\equiv g_{0,i}$ at each site.
Again, in order to
get the periodicity of the chain, one more constraint has to be
imposed
\be
\tau=\sum_{i=1}^{n} \Delta_{i}
\ee
where $f_{n,1}=f_{1,n}\equiv g_{0,n}exp(\Delta_{n})$.
This leads to the potential
\be
V(\phi_{ij})=\sum_{j=1}^{n-1}g_{0,i}^2e^{\phi_{i+1}-\phi_i}+
g_{0,n}^2e^{\phi_{1}-\phi_n}
\ee
However, in this simplest case, we get the same periodic Toda chain, since
the difference of the coupling constants in different sites can be removed by
shifts of the Toda coordinates.

However, there are less trivial limits given
by the choice
\be
f_{i+j,i}=g_{0,j}\exp(\Delta_j)
\ee
that is to some extent complimentary to (\ref{7.1}).
This choice, with specially adjusted $\Delta_i$'s and the proper
periodicity condition, leads to a ``more filled" Lax operator than Toda one,
with several non-zero diagonals so that more pairs $\phi_{i},\phi_{k}$
interact.  In fact, this Lax operator is similar to (\ref{hh2}) and, in the
most general situation, is ``completely filled". This generic system,
certainly, gives rise to the system that differ from the Toda one but the
number of free parameters still can not exceed $n$ ($n-1$ in the $SL(n)$
case).

Turn now to the brane picture behind the described limits. Let us begin with
the rough ``perturbative" configuration. Remind that for the single scale
$N=2$ theory one has a pair of NS5 brane with $n$ D4 branes between them.
The distance in the $x^{6}$ direction is identified with $\frac{1}{g^{2}}$.
Therefore, in order to reproduce the above consideration, one would
order the D0-branes along the
$x^{6}$-coordinate in the simple Toda case but in the general case the
D0-branes are to be assigned with arbitrary positions along
this direction. This provides a set of new scales in field theory.
The corresponding multiscale low-energy theory is broken,
apart from the usual Higgs mechanism of the gauge $SU(n)$
symmetry breaking, by the explicit gauge symmetry breaking at the regulator
scale. Due to the asymptotical freedom of the theory and the conformal
anomaly, this ultraviolet breaking is lifted down to the low energy sector.
In the Lagrangian terms this means that we break $N=4$ theory down to $N=2$
adding more complicated regulators with the set of mass parameters that
manifestly break the gauge symmetry $\Tr\left[M(\Phi_1^2+\Phi_2^2)\right]$,
where $M$ is a mass regulator matrix.

Another interpretation of the regulator scales is related to the
recent proposal \cite{nikita3} that the UV
cut-offs can be identified with the momenta $p_{11}$ in M theory.
Therefore, we can describe our multiscale
case as corresponding to the nondegenerated set of momenta along the eleventh
dimension. This implies the possibility of nontrivial brane scattering
processes with the $p_{11}$ transfer \cite{pp} which should has much to do
with the interpretation of Whitham dynamics as a scattering process \cite{Gor}.

Certainly, entering the several UV scales into the low-energy physics deserves
further studies and we are going to return to this question
in the forthcoming paper \cite{ggm2}.

\section{Discussion on symmetries}
\setcounter{equation}{0}

This section is slightly outside the main line of the paper, but we consider
it important for the future development. In fact, we speculate on what might
be the role of some key notions/properties of integrable systems that have
not been used yet in the SUSY theories/brane framework. In particular, we
briefly discuss the hidden symmetries which are well known as ``non-abelian
symmetries" in the context of the integrable field theories and have some
counterpart in the finite dimensional case. Their role in the classical
dynamics on the spectral curve, is still to be understood but
they certainly are of the great importance at the quantum level\footnote{Let
us fix at the very beginning that all the hidden symmetries under
consideration are the Poisson-Lie ones.}. However, we start with some
discussion of how the finite-dimensional system is embedded into $2d$ one and
only then come to the symmetry issues.

\subsection{Embedding in 2d integrable systems}
In order to see the full (typically infinite dimensional) symmetry
group underlying a finite-dimensional system, one needs to embed it into a
($2d$) infinite dimensional integrable theory. Embedding of such a type is
known, in particular, for the systems of the Calogero type. In fact this is a
heuristic observation that there is a close relation of these systems with
special solutions in 2d integrable field theories \cite{embed}. This
correspondence has no rational explanation yet.

We expect that the embedding of the many-body systems related to the
SUSY/brane theories into $2d$ systems would be very useful in understanding
of the processes of creation/annihilation etc.
To be precise this embedding is formulated as follows
\cite{Kri1}. One should start
with the rational, trigonometric or elliptic solutions to the KP hierarchy
with the $\tau$-function that has a multiplicative form
\be
\tau(x,t)=\prod_{i=1,...,n} \sigma(x-x_{i}(t))
\ee
Now the condition for this $\tau$-function to satisfy the equation
of the KP hierarchy defines the dependence of zeros $x_{i}$ on all
the times of the hierarchy. This dependence is determined by the dynamics
given by the Hamiltonians of the many-body Calogero system, that is to say,
by the traces of the Calogero Lax operator corresponding to the, accordingly,
rational, cylinder or elliptic geometry. The number of factors in the
$\tau$-function coincides with the number of particles in the Calogero system.
Two more requirements are to be added -- the Calogero coupling constant is
to be fixed and is not allowed to take arbitrary values and the evolution has
to be started with the specific ``locus" particle configuration with zero
particle momenta. This correspondence holds for the generalization of
the KP system to the $2d$ Toda lattice hierarchy, the dynamics of the specific
solutions to the $2d$ Toda being governed by Hamiltonians of the Ruijsenaars
finite-dimensional integrable system \cite{kriza}.

Let us remark that according to \cite{embed} one can establish an analogous
correspondence between the Calogero and KdV dynamics. In this case,
the whole procedure can be naturally performed in two steps. First, one fix
the system of $n$ coinciding particles with the total charge $n$, which
further decays (in accordance with the KdV dynamics) into the zero momenta
``locus" configuration that evolves with the Calogero Hamiltonians. This
locus configuration is not arbitrary but possesses $\Z_n$-symmetry.
This
procedure can be also interpreted in brane terms, with the Calogero
degrees of freedom regarded as D0 coordinates. Then, the initial
configuration is interpreted as a bound state of $n$ D0-branes which
KdV-evolve to the stable ``locus" configuration which then evolves with the
Calogero Hamiltonians.

Within the described picture, the possible role that
$2d$ integrable systems like KP or Toda could play looks very
promising. Indeed, we have seen that the dynamics of the
finite number of D0-branes corresponds to the evolution of the very
special solution of $2d$ theory. Therefore, it seems unavoidable
that the general $2d$ dynamics is described by the nontrivial
D0 distribution along the $x^{6}$ direction. Thus, the integrable system
to some extend looks as a second quantized ``brane or string"
theory. Note
that the KdV dynamical variable u(x) now can be treated as the
$x^{6}$-component of the stress energy tensor so that
the expression for the simplest (rational) solution
$$
u(x)=\sum_{i} \frac{1}{(x-x_{i}(t))^{2}}
$$
in D0 brane terms looks quite satisfactory.

This second quantized picture can be pushed even further. To this end,
let us note that embedding different Calogero systems into the same KdV
(or, more generally, KP/$2d$ Toda lattice) hierarchy plays a unifying role.
Indeed, solutions of the Calogero systems with different number of particles,
that is to say, with different gauge group ranks, being considered as
solutions to the same KdV hierarchy are related by a B\"acklund
transformation that can be realized by the (fermionic) vertex operators
\cite{DJKM}. Thus, these vertex operators are associated with the operator of
the brane creation providing the change of the gauge group rank. This means
that the (sub)space of the solutions to the larger hierarchy can be chosen
as a configuration space of all the vacua/string theories. In its particular
realization, this space can be described as an infinite dimensional
Grassmannian \cite{SegWil} and reminds of the analogous construction known
from the perturbative string physics \cite{MorKn}.

In the rest part of this section we discuss the issues of symmetries in
integrable systems that can be easier and more naturally formulated for the
quantum systems. Although so far we have met only the classical integrable
systems, below we follow the general approach equally applicable to both
quantum and classical integrable systems. Among other advantages, this will
make quantization of the construction an immediate thing.

Let us give some general comments concerning the quantum picture.
From the consideration above it is clear that
the integrable dynamics would proceed on the instantonic moduli space
so  all the ingredients of the quantum picture should acquire the meaning
of the appropriate objects on this moduli space. Note that in the course of
quantization, i.e. constructing the wave function, in the present ``Hitchin
treatment" we need to neglect the Whitham variables taking them as just
parameters. This corresponds to the Born-Oppenheimer approximation. It is
just in this approximation the whole wave function reflects the structure of
the instanton moduli space.

As we have argued, instead of the infinite dimensional instanton moduli
space, we work with the finite number of effective variables which substitute
the infinite instanton sums. The wave functions seem to be related to the
cohomology of the moduli space while the quantum partition function would
play the role of the generation function for the intersection numbers in
cohomologies.

Asymptotics of the wave function gives the $S$-matrix for the effective
degrees of freedom -- D0 branes. The integrability and the presence
of the additional conservation laws promise the exactness of the $S$-matrix
as well as its pure pole structure. At the same time, in order to derive
the form-factors (in our context, it corresponds to the form-factor of the
compound state of multiple D0 branes),
one needs to use the hidden non-abelian symmetries which
we are going to discuss now.

\subsection{B\"acklund transformation}

The remnant of the infinite dimensional non-abelian
symmetry in the Toda system is the B\"acklund transformation.
It is known that the nontrivial
B\"acklund transformation for the periodic Toda chain
corresponds to the transition between two equivalent
sets of variables $x_{2n}$ and $x_{2n+1}$.
At the quantum level, it is convenient to consider the wave
function in the action-angle representation which is
the solution to the Baxter equation
\be
Q(\lambda +i)+\Lambda_{QCD}^{2n_{c}}Q(\lambda -i)=Q(\lambda)TrT(\lambda)
\ee
where $T(\lambda)$ is the transfer matrix of the Toda chain and $\lambda$
is the spectral parameter. This equation actually is the quantum counterpart
of the spectral curve equation considered as the operator acting on the wave
function in the separated variables \cite{Sklyanin}\footnote{In the
classical limit, the functions ${Q(\lambda+i)\over Q(\lambda)}$ and
${Q(\lambda-i)\over Q(\lambda)}$ can be identified with $w$ and
${1\over w}$ respectively.}.

The key point is that the function $Q$ turns out to be the
generating function for the
canonical B\"acklund transformation. Namely,
it can be related \cite{back} to the kernel of
the operator acting between two set of variables $q_{i},\bar{q_{i}}$.
Explicit form of the integral kernel of $Q(\lambda)$ looks as follows
\be
Q_{\lambda}(q|\bar{q})=\prod_{j} W_{\lambda}(\bar{q_{j}}-q_{j})
\bar{W_{\lambda}}(q_{j}-\bar{q_{j+1}})
\ee
where
\be
W_{\lambda}(q)=exp(i\lambda q-e^{q}) \\
\bar{W_{\lambda}}(q)=exp(-i\lambda q-e^{q})
\ee
Logarithms of the matrix elements of Q are the
generating functions for the canonical B\"acklund transformation.
This is in agreement with the general viewpoint: $Q$ is actually
related to the Toda $\tau$-function while the $\tau$-function
is the generating function for some canonical transformations.

\subsection{Semiclassical Yangian type symmetry}

Let us turn now to the analogous symmetry in the
integrable system corresponding to the $N=2$ SQCD -- spin chain \cite{be}.
It is known that the spin chains enjoy two different sets
of the conservation laws -- Abelian and non-Abelian ones (they are local and
non-local correspondingly). The charges of the both sets are conserved but,
because of the non-abelian nature of the second set, one can not diagonalize
the Yangian charges simultaneously. The explicit formulae for the Yangian
structure can be found in \cite{be}. Hereafter, we consider the $SL(2)$ spin
chain, it is homogeneous and the Casimirs at all sites are the same. This is
equivalent to the coincidence of all masses of the fundamental matter in
the $N=2$ SQCD.

We now emphasize that the knowledge of charges is equivalent to the
data of the monodromy matrix $T$
(we use here the unit normalization of the determinant):
\be
T(\lambda)=\pmatrix{A(\lambda ) B(\lambda ) \cr C(\lambda )  D( \lambda ) }
\qquad;\qquad \det\ T(\lambda) = AD-BC=1
\ee
Yangian charges can be expressed in terms of the monodromy matrix as follows
\be
{\cal P}_{ij}(\lambda) = \delta_{ij}+ \sum_{n=0}^\infty \lambda^{-n-1}
{\cal P}^n_{ij} =
\half tr( T \sigma_i T^{-1} \sigma_j ) \non
{\cal P}_i(\lambda) = \sum_{n=0}^\infty \lambda^{-n-1} {\cal P}^n_i =
\ep_{ijk} {\cal P}_{jk}(\lambda) \ \ \ \ i=1,2,3
\ee
Therefore, the generators ${\cal P}_{ij}(\lambda)$ and ${\cal P}_i(\lambda)$
are quadratic
functions of the matrix elements of $T(\lambda)$.

It is also possible to express $T(\lambda)$ in terms
of the generating functions ${\cal P}_i(\lambda)$:
\be
T(\lambda)=
{\inv{2}}W(\lambda){\bf 1} -{\frac{i}{2}} W^{-1}(\lambda)
\sum_i {\cal P}_i(\lambda) \sigma_i
\ee
with $ W(\lambda)= \sqrt{ 2 + \sqrt{4-\vec{{\cal P}}^2(\lambda)} }$.

Thus, the charges ${\cal P}_i(\lambda)$ contain the same amount of
information as the monodromy matrix.
There are also the following relations
between the ${\cal P}_i(\lambda)$ and the matrix elements of $T(\la)$:
\be
{\cal P}_+(\lambda) = {\cal P}_1(\lambda) +i {\cal P}_2(\lambda) = 2i
W(\lambda)\  C(\lambda) \label{nlc1} \\
{\cal P}_-(\lambda) = {\cal P}_1(\lambda) -i {\cal P}_2(\lambda) = 2i
W(\lambda)\  B(\lambda) \label{nlc2}\\
\Tr T(\lambda)=W(\lambda)=W\left[\vec{{\cal P}}^2(\lambda)\right]
\ee
Therefore, $\vec{{\cal P}}^2(\lambda)$ is also a generating
function for (non-local) commuting quantities.
\par

For the completeness, let us manifestly write down the Poisson brackets of
semiclassical Yangian generators
\be
\{ {\cal P}_i^0,{\cal P}_j^0\} = 4 \ep_{ijk} {\cal P}^0_k \non
\{ {\cal P}_i^0,{\cal P}_j^1\} = 4 \ep_{ijk} {\cal P}^1_k \label{CIxi}\\
\{{\cal P}^1_i,\{{\cal P}^1_j,{\cal P}^0_k\}\} -\{{\cal P}^0_i,\{{\cal P}^1_j,
{\cal P}^1_k\}\} =
A_{ijk}^{lmn} {\cal P}^0_l {\cal P}^0_m {\cal P}^0_n \non
\{\{{\cal P}^1_i,{\cal P}^1_j\},\{{\cal P}^0_k,{\cal P}^1_l\} \}
+\{\{{\cal P}^1_k,{\cal P}^1_l\},\{{\cal P}^0_i,{\cal P}^1_j\}\} =
8( A_{ija}^{mnp}\ep_{kla} + A^{mnp}_{kla} \ep_{ija}) {\cal P}^0_m {\cal P}^0_n
{\cal P}^1_p
\nonumber
\ee
with $A_{ijk}^{lmn}= {\textstyle{2\over
3}}\ep_{ila}\ep_{jmb}\ep_{knc}\ep^{abc}$.

One sees that these relations form a deformation of those defining
the Borel part of the $su(2)$ loop algebra.
All the non-local
charges ${\cal P}_{ij}^n$ can be expressed in terms of the Poisson brackets
between the two first charges ${\cal P}_i^0$ and ${\cal P}_i^1$. Therefore,
the whole Poisson algebra of the symmetries is generated by these two
charges. They are nothing but the corresponding Chevalley generators.

Finally, let us describe how the monodromy matrix generates
the transformations of the spin variable on the $k$-th site
which corresponds in the brane language to the action
on the $k$-th (D6+D4) brane.
By an explicit computation of the Poisson brackets between
the monodromy matrix and the spin variables, one can show
that the variation $\delta_v^n S(k)$ of the spin variables
reads as:
\be
\delta_i^n S(k)\equiv\left\{{\cal P}^n_i,S(k)\right\}=
\ i\oint {{d\lambda }\over{2i\pi}} \lambda^n\
\Tr_1\BL\ (\sigma_i T^{-1}(\lambda)\otimes 1)\Bigl\{
T(\lambda)\otimes 1 , 1\otimes S(k) \Bigr\}\ \BR \label{CIxvi}
\ee
Here $\Tr_1$ denotes the trace over the first space in the tensor product and
$v_i$ is the parameter of the transformation. This formula describes
the transformations $S(k)\to \delta^n_vS(k)$ and indicates that
$T(\lambda)$ also generates the non-Abelian symmetry.

Let us comment now the possible meaning of the non-abelian symmetry within
the brane approach. Usually this symmetry is important at the quantum level
providing the set of restrictions onto the $S$-matrix. In our context it would
mean that the symmetry restricts the brane scattering.
Note that the Yangian charges are essentially
non-local. In the SQCD context this would correspond to mixing the different
flavors (that are associated with different sites of chain) and presumably
might be relevant for the barionic branch of the theory.

\section{Conclusion}

In the present paper we discussed effective $N=2$ SUSY
field theories, corresponding integrable systems, monopoles and
their stringy analogues. The equivalences were established at low energies
in field theories as well as in string theory models. One can try
to go further and ask whether such connections persist
if we consider higher derivatives in field theory corresponding to
non-lowest excitations in string theory. The answer seems to be yes,
and there is  natural candidates in the world of integrable systems
for the counterpart of higher-derivative terms. They are related to the higher
times and introducing two-dimensional integrable systems. We are going to
elaborate this point in the subsequent publication \cite{ggm2}. We also
postponed the discussion of elliptic models (and their degenerations)
as well as five- and six-dimensional field theories (XXZ and XYZ
spin chains), which constitute the two last rows in Fig.1. The new
examples confirm the validity of the general approach and suggest
the proper way to include additional degrees of
freedom. Moreover, it provides the possibility to get completely
new viewpoints in the field theory setup which we have
seen in section 7. Nevertheless, despite many
supporting arguments including the identification of
branes as the proper degrees of freedom, the clear
{\it physical} answer
to the question: ``Why the low-energy SYM is governed
by integrable system?" is still missed.

The answer to this question requires the knowledge of the exact
derivation of the effective degrees of freedom in the integrable
systems from the instantonic sums. We have argued that
somehow the instantons are summed into the finite number of degrees
of freedom. However, the proper
derivation and explanation of this phenomena from the path
integral of SYM is still lacking. The answer would imply a
nontrivial localization while integrating over the instanton moduli
space.

So far only the tiny part of the ``integrability world"
has been recognized in the supersymmetric YM theories.
Say, the issues of underlying symmetries especially in connection with
the quantization problem as well
as interpretation of the non-Abelian ``spectrum generating" symmetries
in the brane language have been out of mainstream. We are planning to
discuss this points elsewhere.

Another important point about the integrable treatment
is to interpret the wave functions
(which would be interpreted as the correlators in a
topological theory) that should provide the important information
about the vacuum configurations. The wave functions actually
serve as generating functions for the canonical
transformations. It is interesting to identify
these canonical transformations in the brane language.

Let us say some words about the possible role of integrability
in $N=1$ SUSY theory. It was recently shown \cite{Ooguri,QCD} that
the perturbative
spectral curve in such a theory is a sphere with marked points which suggests
that we leave the family of the ``finite-gap" solutions. But
it can be not the whole story. Indeed,
the branching points on the spectral curves come
from the Coulomb branch moduli and so $\Lambda_{QCD}$ does. The
Coulomb moduli degenerate to the set of points which
means that from the $N=2$ SUSY point of view we can no longer vary the
integrals of motion. But the potential cut in the strong
coupling region still persists due to $\Lambda_{QCD}$ so the
one-gap solutions are potentially possible.

Recently another object from the integrability cuisine was
also discussed in the context of $N=1$ theory, that is, the
``creek equation" from \cite{shifman} that gives
the profile of the BPS objects.
This equation is a direct counterpart of the
Nahm equation treatment above. So the would be integrable
system in $N=1$ context corresponds to the dynamics
of several interacting BPS states .

We have seen some similarities with the ``confining string"
scenario above. Indeed, one of the main points of that scenario is
the presence of ``the monopole ring" in the vacuum state,
the appearance of the effective ``confining string"
to substitute the infinite number of instanton and the
appearance of some auxiliary surface whose genus corresponds
to the rank of the gauge group. One immediately recognizes these
ingredients in the integrability approach but we do not know
any real statements behind this observation.

Note that
the picture considered in the paper has, in fact, some solid state
analogies. There are many examples in the solid state physics when the mass
gap in the spectrum of the quasiparticles vanishes on (hyper)surfaces of
different dimensions in momentum space. Just
these ``defects" (``monopoles") in the momentum space are counterparts of
branes in our approach. The very appearance of the (topologically nontrivial)
hypersurfaces in the theories with anomalies has a simple interpretation.
Indeed, the presence of the anomaly (for instance, the conformal anomaly in
the SYM theory) implies the nontrivial level crossing that can be seen within
the standard Berry phase treatment. In our consideration the mass of
regulator (entering the low-energy sector due to the conformal anomaly) gives
rise to the ``monopole field" in the space of fields in agreement with this
interpretation.

\bigskip

We thank A.Kapustin, N.Nekrasov, S.Kharchev, I.Krichever, A.Levin,
A.Marshakov, A.Morozov, M.Ol\-sha\-net\-sky, I.Polyubin,
A.Vainshtein, Z.Yin and A.Zabrodin for the useful discussions.
A.G. thanks H.Leut\-wy\-ler for the hospitality in the Institute of
Theoretical Physics at Bern University, where the part of the work was done.
S.G. is grateful to the organizing committee of summer school in Cargese
for hospitality, where the part of this work was done. The work of
A.G. was partially supported by grants CRDF RP2-132, INTAS-93-0273,
the work of S.G.  -- by grant RFBR-96-15-96939, that
of A.M.  -- by grants RFBR-96-02-16210(a), INTAS-97-1038.

\section{Appendix. $SL(3)$ chain}
\setcounter{equation}{0}
\def\theequation{A.\arabic{equation}}
In this Appendix we consider the simple case of the $SL(3)$ chain to
give explicit examples of some formulas of sect.\ref{magnet}.

\paragraph{$SL(3)$ spin chain.}
The Lax operator is given by the formula
\be\label{Lax3}
L_i=\left(
\begin{array}{ccc}
S_{0,i}^{(1)}+\lambda+\lambda_i&S_{-,i}^{(1)}&S_{-,i}^{(12)}\\
S_{+,i}^{(1)}&S_{0,i}^{(2)}+\lambda+\lambda_i&S_{-,i}^{(2)}\\
S_{+,i}^{(12)}&S_{+,i}^{(2)}&-S_{0,i}^{(1)}-S_{0,i}^{(2)}+\lambda+\lambda_i
\end{array}
\right)
\ee
The spectral curve in this case has the form
\be\label{sc3}
w^3+\Tr T(\lambda) w^2 + \sum_i \det M_i(\lambda) w+\det T(\lambda)=0
\ee
where $T(\lambda)$ is, as before, the transfer matrix and $M_i(\lambda)$ is
matrix obtained from the transfer matrix by removing the $i$-th column
and $i$-th row.
In this reference integrable system, $J_1=1$ and $J_2=\det
T(\lambda)=\prod_i\det L_i(\lambda)$.  To determine mass spectrum we need to
calculate $\det L_i(\lambda)$. It has the form
\be\label{detL}
\det L_i(\lambda)=(\lambda+\lambda_i)^3-C^{(i)}_2(\lambda+\lambda_i)-
C^{(i)}_3
\ee
where Casimir functions are
\be
C_2^{(i)}=S_{+,i}^{(1)}S_{-,i}^{(1)}+S_{+,i}^{(2)}S_{-,i}^{(2)}+
S_{+,i}^{(12)}S_{-,i}^{(12)}+\left(S_{0,i}^{(1)}\right)^2+
\left(S_{0,i}^{(2)}\right)^2+S_{0,i}^{(1)}S_{0,i}^{(2)}\\
C_3^{(i)}=
S_{-,i}^{(12)}S_{+,i}^{(12)}S_{0,i}^{(2)}
+S_{-,i}^{(2)}S_{+,i}^{(2)}S_{0,i}^{(1)}
-S_{+,i}^{(1)}S_{+,i}^{(2)}S_{-,i}^{(12)}
-S_{-,i}^{(1)}S_{-,i}^{(2)}S_{+,i}^{(12)}+\\
+S_{0,i}^{(1)}S_{0,i}^{(2)}\left(S_{0,i}^{(1)}+S_{0,i}^{(2)}\right)
-S_{-,i}^{(1)}S_{+,i}^{(1)}\left(S_{0,i}^{(1)}+S_{0,i}^{(2)}\right)
\ee
that can be manifestly checked using the Poisson brackets
\be\label{pbsl3}
\{S_+^{(1)},S_-^{(1)}\}=S_0^{(2)}-S_0^{(1)},\ \ \ \
\{S_+^{(2)},S_-^{(2)}\}=-S_0^{(1)}-2S_0^{(2)},\ \ \ \
\{S_{\pm}^{(1)},S_0^{(1)}\}=\pm S_{\pm}^{(1)},\\
\{S_{\pm}^{(1)},S_0^{(2)}\}=\mp S_{\pm}^{(1)},\ \ \ \
\{S_{\pm}^{(2)},S_0^{(1)}\}=0,\ \ \ \
\{S_{\pm}^{(2)},S_0^{(2)}\}=S_{\pm}^{(2)},
\ee
and defining relations
\be\label{pbsl3'}
\{S_{\pm}^{(1)},S_{\pm}^{(2)}\}=\mp S_{\pm}^{(12)}
\ee
so that
\be\label{pbsl3''}
\{S_-^{(1)},S_+^{(12)}\}=-S_+^{(2)},\ \ \ \
\{S_-^{(2)},S_+^{(12)}\}=S_+^{(1)},\ \ \ \
\{S_+^{(1)},S_-^{(12)}\}=S_-^{(2)},\ \ \ \
\{S_+^{(2)},S_-^{(12)}\}=-S_-^{(1)},\\
\{S_{\pm}^{(12)},S_0^{(1)}\}=\pm S_{\pm}^{(12)},\ \ \ \
\{S_{\pm}^{(12)},S_0^{(2)}\}=0,\ \ \ \
\{S_-^{(12)},S_+^{(12)}\}=2S_0^{(1)}+S_0^{(2)}
\ee
Generally, the determinant (\ref{detL}) has three distinct roots
(masses), since we deal with
the reference system, that is to say, $J_1(\lambda)=1$. In
order to obtain non-unit $J_1(\lambda)$, one needs to consider the case of,
two coinciding roots. This condition implies specialization of the general
orbit
\be\label{C2=C3}
4C_2^3=27C_3^2
\ee
and the coinciding masses are equal to
\be
m_c=-\lambda_i\pm\sqrt{{C_2\over 3}}
\ee
Then, the factor $(\lambda-m_c)^2$ is nothing but $J_1(\lambda)$
($J_1(\lambda)$ of the general form can be obtained by degenerating at a set
of sites). Now one can check that condition (\ref{C2=C3}) leads to factoring
out the multiplier $(\lambda-m_c)$ in the coefficient in front of $w$ in
(\ref{sc3}) to reproduce the correct dependence of the spectral curve on
$J_1(\lambda)$. This coefficient is equal to
\be\label{1}
T_{11}T_{22}+T_{11}T_{33}+T_{22}T_{33}-T_{12}T_{21}-T_{13}T_{31}-
T_{23}T_{32}
\ee
For the definiteness, let us consider the degeneration at the first site.
Note that (\ref{1}) can be rewritten as
\be\label{2}
\left(L_{1,11}L_{1,22}-L_{1,12}L_{1,21}\right)
\left({\hat T}_{11}{\hat T}_{22}-{\hat T}_{12}{\hat T}_{21}\right)+\\+
\left(L_{1,11}L_{1,33}-L_{1,13}L_{1,31}\right)
\left({\hat T}_{11}{\hat T}_{33}-{\hat T}_{13}{\hat T}_{31}\right)+\\+
\left(L_{1,22}L_{1,33}-L_{1,23}L_{1,32}\right)
\left({\hat T}_{22}{\hat T}_{33}-{\hat T}_{23}{\hat T}_{32}\right)
\ee
where ${\hat T}$ is the transfer matrix without the first site and
$L_{1,ij}$ is $(i,j)$-matrix element of $L_1$. This
expression can be rewritten as follows
\be
\left(\lambda+\lambda_i+{1\over 2}\left(S_{0,1}^{(1)}+S_{0,1}^{(2)}\right)
+{1\over 2}\sqrt{\left(S_{0,1}^{(1)}-S_{0,1}^{(2)}\right)^2+
4S_{-,1}^{(1)}S_{+,1}^{(1)}}\right)\times\\\times
\left(\lambda+\lambda_i+{1\over 2}\left(S_{0,1}^{(1)}+S_{0,1}^{(2)}\right)
-{1\over 2}\sqrt{\left(S_{0,1}^{(1)}-S_{0,1}^{(2)}\right)^2+
4S_{-,1}^{(1)}S_{+,1}^{(1)}}\right)\times\\\times
\left({\hat T}_{11}{\hat T}_{22}-{\hat T}_{12}{\hat T}_{21}\right)+\\
+
\left(\lambda+\lambda_i-{1\over 2}S_{0,1}^{(2)}
+{1\over 2}\sqrt{\left(S_{0,1}^{(2)}+2S_{0,1}^{(1)}\right)^2+
4S_{-,1}^{(12)}S_{+,1}^{(12)}}\right)\times\\\times
\left(\lambda+\lambda_i-{1\over 2}S_{0,1}^{(2)}
-{1\over 2}\sqrt{\left(S_{0,1}^{(2)}+2S_{0,1}^{(1)}\right)^2+
4S_{-,1}^{(12)}S_{+,1}^{(12)}}\right)\times\\\times
\left({\hat T}_{11}{\hat T}_{33}-{\hat T}_{13}{\hat T}_{31}\right)+\\+
\left(\lambda+\lambda_i+{1\over 2}S_{0,1}^{(1)}
+{1\over 2}\sqrt{\left(S_{0,1}^{(1)}+2S_{0,1}^{(2)}\right)^2+
4S_{-,1}^{(2)}S_{+,1}^{(2)}}\right)\times\\\times
\left(\lambda+\lambda_i+{1\over 2}S_{0,1}^{(1)}
-{1\over 2}\sqrt{\left(S_{0,1}^{(1)}+2S_{0,1}^{(2)}\right)^2+
4S_{-,1}^{(2)}S_{+,1}^{(2)}}\right)\times\\\times
\left({\hat T}_{22}{\hat T}_{33}-{\hat T}_{23}{\hat T}_{32}\right)
\ee
Now by the straightforward calculation one can check that each of three
terms in this formula is proportional to $(\lambda-m_c)$ provided the
condition (\ref{C2=C3}) is satisfied.

\paragraph{$SL(3)$-Toda-like system.}
Now let us consider the ``limiting" degenerations of the Lax operator
(\ref{Lax3}) analogous to the Toda system in the $SL(2)$ case.
We follow $SL(2)$ case and multiply the Lax operator (\ref{Lax3}) by the
constant
diagonal matrix $U$. In this case, there are two possibilities. First
possibility is to choose
\be
U=\left(
\begin{array}{ccc}
1&0&0\\
0&\alpha&0\\
0&0&1
\end{array}
\right)
\ee
and redefine $S_+^{(1)}\to{1\over\alpha}S_+^{(1)}$,
$S_-^{(2)}\to{1\over\alpha}S_-^{(2)}$. Now, bringing $\alpha$ to zero, one
obtains the Lax operator
(hereafter, we omit inhomogeneity, since it does not effect the result)
\be\label{Lax3d1}
L=\left(
\begin{array}{ccc}
S_{0}^{(1)}+\lambda&S_{-}^{(1)}&S_{-}^{(12)}\\
S_{+}^{(1)}&0&S_{-}^{(2)}\\
S_{+}^{(12)}&S_{+}^{(2)}&-S_{0}^{(1)}-S_{0}^{(2)}+\lambda
\end{array}\right)
\ee
Then, instead of (\ref{pbsl3}), we get
\be\label{pbsl3d}
\{S_+^{(1)},S_-^{(1)}\}=\{S_+^{(2)},S_-^{(2)}\}=0,\ \ \ \
\{S_{\pm}^{(1)},S_0^{(1)}\}=\pm S_{\pm}^{(1)},\ \ \ \
\{S_{\pm}^{(1)},S_0^{(2)}\}=\mp S_{\pm}^{(1)},\\
\{S_{\pm}^{(2)},S_0^{(1)}\}=0,\ \ \ \
\{S_{\pm}^{(2)},S_0^{(2)}\}=\pm S_{\pm}^{(2)}
\ee
and instead of (\ref{pbsl3'}) --
\be\label{pbsl3'd}
\{S_{\pm}^{(1)},S_{\pm}^{(2)}\}=0
\ee
while (\ref{pbsl3''}) remain unchanged (now, however, one can not use
the defining relations (\ref{pbsl3'}) to generate all other (\ref{pbsl3''})
but they should be given independently).

At the next step, the second Casimir function
of the algebra (\ref{pbsl3d})-(\ref{pbsl3'd})
$S_{+}^{(1)}S_{-}^{(1)}+S_{+}^{(2)}S_{-}^{(2)}$ should be put equal to zero.
Then the determinant of the Lax operator (\ref{Lax3d1}) is equal to the third
Casimir function
\be
S_{+}^{(1)}S_{+}^{(2)}S_{-}^{(12)}
+S_{-}^{(1)}S_{-}^{(2)}S_{+}^{(12)}
+S_{-}^{(1)}S_{+}^{(1)}\left(2S_{0}^{(1)}+S_{0}^{(2)}\right)
\ee
that can be chosen equal to unity.

The second possibility of degenerations of the Lax operator
(\ref{Lax3}) looks slightly simpler having less non-trivial entries.
It can be achieved by
multiplying the Lax operator by the matrix $U$ of the form
\be
U=\left(
\begin{array}{ccc}
1&0&0\\
0&\alpha&0\\
0&0&\alpha
\end{array}
\right)
\ee
This form of the matrix allows one to redefine, apart from
$S_+^{(1)}\to{1\over\alpha}S_+^{(1)}$,
$S_{\pm}^{(2)}\to{1\over\alpha}S_{\pm}^{(2)}$
$S_+^{(12)}\to{1\over\alpha}S_+^{(12)}$, also the diagonal element
$S_0^{(2)}\to{1\over\alpha}S_0^{(2)}$ {\it without}  tuning the
inhomogeneity (compare with the $SL(2)$ case). Now taking $\alpha$ to zero,
one obtains the Lax operator
\be\label{Lax3d2}
L=\left(
\begin{array}{ccc}
\lambda+S_0^{(1)}&S_-^{(1)}&S_-^{(12)}\\
S_+^{(1)}&S_0^{(2)}&S_-^{(2)}\\
S_+^{(12)}&S_+^{(2)}&-S_0^{(2)}
\end{array}\right)
\ee
and algebra of the Poisson brackets
\be\label{pb3d2}
\{S_+^{(1)},S_-^{(1)}\}=S_0^{(2)},\ \ \ \
\{S_+^{(2)},S_-^{(2)}\}=0,\ \ \ \
\{S_{\pm}^{(1)},S_0^{(1)}\}=\pm S_{\pm}^{(1)},\ \ \ \
\{S_{\pm}^{(1)},S_0^{(2)}\}=\{S_{\pm}^{(2)},S_0^{(1)}\}=0,\\
\{S_{\pm}^{(2)},S_0^{(2)}\}=0,\ \ \ \
\{S_{\pm}^{(1)},S_{\pm}^{(2)}\}=0,\ \ \ \
\{S_-^{(1)},S_+^{(12)}\}=-S_+^{(2)},\ \ \ \
\{S_-^{(2)},S_+^{(12)}\}=\{S_+^{(2)},S_-^{(12)}\}=0,\\
\{S_+^{(1)},S_-^{(12)}\}=S_-^{(2)},\ \ \ \
\{S_{\pm}^{(12)},S_0^{(1)}\}=\pm S_{\pm}^{(12)},\ \ \ \
\{S_{\pm}^{(12)},S_0^{(2)}\}=0,\ \ \ \
\{S_-^{(12)},S_+^{(12)}\}=2S_0^{(1)}+S_0^{(2)}
\ee
One can easily check that $S_0^{(2)}$ and $S_{\pm}^{(2)}$ are
the Casimir functions of the algebra (\ref{pb3d2}). In order to get constant
determinant of the Lax operator (\ref{Lax3d2}) we put $S_{\pm}^{(2)}=\pm 1$
and $S_0^{(2)}=1$.Finally this leads us  to the Lax operator
\be\label{Lax3d2'}
L=\left(
\begin{array}{ccc}
\lambda+S_0^{(1)}&S_-^{(1)}&S_-^{(12)}\\
S_+^{(1)}&1&-1\\
S_+^{(12)}&1&-1
\end{array}\right)
\ee
and the Poisson bracket algebra
\be\label{116}
\{S_+^{(1)},S_-^{(1)}\}=1,\ \ \ \
\{S_{\pm}^{(1)},S_{\mp}^{(12)}\}=-1,\ \ \ \
\{S_{\pm}^{(1)},S_0^{(1)}\}=\pm S_{\pm}^{(1)},\ \ \ \
\{S_{\pm}^{(12)},S_0^{(1)}\}=\pm S_{\pm}^{(12)}
\ee
The determinant of the Lax operator (\ref{Lax3d2'}) is equal to the Casimir
function
\be\label{117}
C=S_+^{(1)}S_-^{(1)}+S_+^{(1)}S_-^{(12)}-S_+^{(12)}S_-^{(12)}-
S_-^{(1)}S_+^{(12)}
\ee
and also can be put equal to unity.

Let us note that (\ref{Lax3d2'}) describes a slightly specified case of the
general situation, since the constant sub-matrix
$\pmatrix{1&-1\cr 1&-1}$ in the Lax operator is chosen to be
traceless. It can be avoided by a proper rescaling of $S_0^{(1)}$ and
inhomogeneity. This particular choice is one that admits the $n\times n$
representation. As any $n\times n$ representation for the $SL(p)$, $p>2$
magnet, it describes a degenerated spectral curve as compared to the general
pure gauge theory (see s.5.3). Let us trace out how it happens in this
concrete case.

First, one can observe that the system of the Poisson brackets (\ref{116})
has Casimir functions $S_{\pm}^{(1)}\mp S_{\pm}^{(12)}$. Taking into account
that we put Casimir function (\ref{117}) to be unit, we can choose
$S_{\pm}^{(12)}=\pm S_{\pm}^{(12)}\mp c_{\pm}$, $c_+c_-=1$. The simplest
choice is $c_+=c_-=1$. Then, literally following s.5.1, one gets the $n\times
n$ representation:
\begin{equation}
\label{LaxTCL}
{\cal L}(w) =
\left(\begin{array}{ccccccc}
S_{0,1}^{(1)} & 1 & 0 & &&{1\over w}&{S_{+,n}^{(1)}+S_{-,n}^{(1)}-1
\over w}\\
S_{+,2}^{(1)}+S_{-,2}^{(1)}-1
& S_{0,2}^{(1)} & 1 && \ldots & 0&{1\over w}\\
1 & S_{+,3}^{(1)}+S_{-,3}^{(1)}-1
 & S_{0,3}^{(1)} &&\ldots & &0 \\
 && & &\ddots & \\
 && \ldots& &&\ddots &\vdots \\
-{w} & 0 & 0& & & &S_{0,n}^{(1)}
\end{array} \right)
\end{equation}
with the following Poisson brackets: $\{S_{+,i}^{(1)},S_{-,j}^{(1)}\}=
\delta_{ij}$, $\{S_{\pm,i}^{(1)},S_{0,j}^{(1)}\}=\pm S_{\pm,i}^{(1)}
\delta_{ij}$ that can be realized by the two harmonic oscillators
$\{p_a,q_b\}=\delta_{ab}$, $a,b=1,2$: $S_+^{(1)}=p_1e^{q_2}$,
$S_-^{(1)}=q_1e^{-q_2}$, $S_0^{(1)}=p_2$.

The Lax operator (\ref{LaxTCL}) leads to the following spectral curve (if
$n>2$)
\be
w+P^{(1)}_n(\lambda)+P^{(2)}_{n-2}(\lambda){1\over w}+{1\over w^2}=0
\ee
In fact, the low diagonal of units in the Lax operator is nothing but the
product of $c_+c_-$. Requiring this product to be zero, we can reproduce the
particular Lax operator of the construction \cite{KP} (see s.5.2) and the
spectral curve takes the form
\be\label{**}
w+P^{(1)}_n(\lambda)+P^{(2)}_{n-2}(\lambda){1\over w}=0
\ee
Physically this corresponds to bringing one of the NS5-branes to infinity
that results to the two NS5-branes remaining with the semi-infinite branes
attached to one of these NS5-branes (cf. (\ref{**}) with (\ref{hren2})).

To conclude this Appendix, we see that there are, indeed, two different
degenerations, one being characterized by the Lax operator (\ref{Lax3d1})
that contains two diagonal elements with the spectral parameter $\lambda$,
and the other one whose Lax operator contains only one diagonal element with
$\lambda$. Let us note that, unlike the $SL(2)$ case, bosonization of the
both $SL(3)$ limiting degenerations looks quite involved.

\end{document}